\documentclass[a4paper]{article}

\usepackage[margin=2.5cm]{geometry} 
\usepackage[frozencache=true,cachedir=.]{minted}

\usepackage{listings}
\usepackage{graphicx}

\usepackage{hyperref}
\hypersetup{
	colorlinks,
	citecolor=blue,
	breaklinks,
	urlcolor=blue, 
	linkcolor=blue, 
	pdfproducer={LaTeX},
	pdfcreator={pdflatex}
}

\usepackage{amssymb}
\usepackage{epstopdf}

\usepackage{pythonhighlight}
\usepackage[shortlabels]{enumitem}
\usepackage{xcolor}

\usepackage{xspace}

\usepackage[font=normalsize,labelfont=bf]{caption}
\usepackage[font=normalsize]{subcaption}

\newcommand{\code}[1]{\pyth{#1}\xspace}

\usepackage{amsmath}

\DeclareMathOperator*{\argmin}{arg\,min}

\usepackage{bm}

\usepackage[most]{tcolorbox}
\definecolor{darkspringgreen}{rgb}{0.09, 0.45, 0.27}

\usepackage{algorithm}
\usepackage{algpseudocode}
\algnewcommand{\Inputs}[1]{%
  \State \textbf{Inputs:}
  \Statex \hspace*{\algorithmicindent}\parbox[t]{.8\linewidth}{\raggedright #1}
}
\algnewcommand{\Initialize}[1]{%
  \State \textbf{Initialize:}
  \Statex \hspace*{\algorithmicindent}\parbox[t]{.8\linewidth}{\raggedright #1}
}

\newcommand{\scalprod}[3]{\left\langle #2,#3\right\rangle_{#1}}

\usepackage{scalerel}
\usepackage{tikz}
\usetikzlibrary{svg.path}

\definecolor{orcidlogocol}{HTML}{A6CE39}
\tikzset{
  orcidlogo/.pic={
    \fill[orcidlogocol] svg{M256,128c0,70.7-57.3,128-128,128C57.3,256,0,198.7,0,128C0,57.3,57.3,0,128,0C198.7,0,256,57.3,256,128z};
    \fill[white] svg{M86.3,186.2H70.9V79.1h15.4v48.4V186.2z}
                 svg{M108.9,79.1h41.6c39.6,0,57,28.3,57,53.6c0,27.5-21.5,53.6-56.8,53.6h-41.8V79.1z M124.3,172.4h24.5c34.9,0,42.9-26.5,42.9-39.7c0-21.5-13.7-39.7-43.7-39.7h-23.7V172.4z}
                 svg{M88.7,56.8c0,5.5-4.5,10.1-10.1,10.1c-5.6,0-10.1-4.6-10.1-10.1c0-5.6,4.5-10.1,10.1-10.1C84.2,46.7,88.7,51.3,88.7,56.8z};
  }
}

\newcommand\orcidicon[1]{\href{https://orcid.org/#1}{\mbox{\scalerel*{
\begin{tikzpicture}[yscale=-1,transform shape]
\pic{orcidlogo};
\end{tikzpicture}
}{|}}}}

\title{Core Imaging Library -- Part II: Multichannel reconstruction for dynamic and spectral tomography}
\author{Evangelos Papoutsellis\, \orcidicon{0000-0002-1820-9916}$^{\tiny{1,2}}$ \and 
Evelina Ametova\, \orcidicon{0000-0002-8867-3001}$^{1,6}$ \and 
Claire Delplancke\, \orcidicon{0000-0001-7483-0419}$^{4}$ \and 
Gemma Fardell\, \orcidicon{0000-0003-2388-5211}$^{2}$ \and 
Jakob S. J\o{}rgensen\, \orcidicon{0000-0001-9114-754X}$^{3,5}$ \and  
Edoardo Pasca\, \orcidicon{0000-0001-6957-2160}$^{2}$ \and 
Martin Turner\, \orcidicon{0000-0003-0117-8049}$^{7}$ \and 
Ryan Warr\, \orcidicon{0000-0002-7904-0560}$^{1}$ \and 
William R. B. Lionheart\, \orcidicon{0000-0003-0971-4678}$^{3}$ \and  Philip J. Withers\, \orcidicon{0000-0002-1946-5647}$^{1}$}

\date{$^{1}$ Henry Royce Institute, Department of Materials, The University of Manchester, Oxford
Road, Manchester, M13 9PL, United Kingdom \\
$^{2}$ Scientific Computing Department, STFC, UKRI, Rutherford Appleton Laboratory,
Harwell Campus, Didcot OX11 0QX, United Kingdom\\
$^{3}$ Department of Mathematics, The University of Manchester, Oxford Road, Manchester,
M13 9PL, United Kingdom\\
$^{4}$ Department of Mathematical Sciences, University of Bath, Claverton Down, Bath, BA2 7AY, United Kingdom\\
$^{5}$ Department of Applied Mathematics and Computer Science, Technical University of
Denmark, Richard Petersens Plads, Building 324, 2800 Kgs. Lyngby, Denmark\\
$^{6}$ Laboratory for Applications of Synchrotron Radiation, Karlsruhe Institute of
Technology, Kaiserstr. 12, 76131, Karlsruhe, Germany\\[2pt]
$^{7}$ Research IT Services, The University of Manchester, Oxford Road, Manchester M13
9PL, United Kingdom\\[5pt]
$^*$Corresponding author: \url{evangelos.papoutsellis@stfc.ac.uk}}

\newcommand{\module}[1]{\textbf{#1}}  
\usepackage{wasysym}
\usepackage{textcomp}
\usepackage{booktabs}

\begin{document}

\maketitle

\begin{abstract}
The newly developed Core Imaging Library (CIL) is a flexible plug and play library for tomographic imaging with a specific focus on iterative reconstruction. CIL provides building blocks for tailored regularised reconstruction algorithms and explicitly supports multichannel tomographic data. In the first part of this two-part publication, we introduced the fundamentals of CIL. This paper focuses on applications of CIL for multichannel data, e.g., dynamic and spectral. We formalise different optimisation problems for colour processing, dynamic and hyperspectral tomography and demonstrate CIL's capabilities for designing state of the art reconstruction methods through case studies and code snapshots.
\end{abstract}




\maketitle



\section{Introduction}

Over recent years in X-ray Computed Tomography (CT) there has been a growing interest in Dynamic CT, and Spectral CT thanks to the technological advancements on detector speed and sensitivity and on multichannel photon-counting detectors (PCDs), as depicted in the EPSRC Tomography roadmap, \cite{survey}. 

In Dynamic CT the aim is to reconstruct a series of images and depict the complete spatiotemporal response of the scanned object. These temporal variations may occur because the composition/structure evolved, e.g. corrosion, or the object is subject to external input, e.g. compression, or the object moved during the scanning process.

In Spectral CT, using a pixelated energy sensitive detector, it is possible to collect $n$ energy specific radiographs, where $n$ is the number of energy channels. As a result for any voxel in the system it is possible to reconstruct the profile of attenuation coefficient as a function of energy, or conversely to create a tomogram corresponding to each energy bin. Since each chemical element has a characteristic attenuation profile this provides a fingerprint of the elements in each voxel. This fingerprint is especially clear for attenuation spectra that includes the energies corresponding to the X-ray absorption edges (K-edges) for the elements concerned, because there is an abrupt change in attenuation on either side of the edge, \cite{Kruger1977}. Moreover, in pulsed neutron imaging data, sharp edges can also be imaged, \cite{Santisteban2002}. In this case the edges, i.e., the Bragg edges, correspond to abrupt increases in the transmitted spectrum, when the energy is below that possible for Bragg diffraction out of the beam for each diffraction peak providing unique fingerprints corresponding to different crystal structures. In general for spectral imaging, because the signal is allocated to a number of energy bins rather than accumulated to give a single image, the energy resolved data acquired usually suffer from low signal-to-noise ratio, acquisition artifacts and angular undersampling making tomographic image reconstruction difficult.

The scope of this paper is to present the capabilities of the Core Imaging Library (CIL) (\url{https://www.ccpi.ac.uk/cil}), releases available at \cite{ZenodoCIL}, of the Collaborative Computational Project in Tomographic imaging (CCPi) for multichannel tomography. It allows one to reconstruct higher quality images and ensure more accurate spatiospectral K-edge identification, see for instance \cite{Ryan, Evelina}, where novel reconstruction methods are introduced for lab-based hyperspectral CT and Neutron Tomography respectively. CIL is an open source, object-oriented library (primarily written in Python) for processing tomographic data. We can  read, load, preprocess, reconstruct, visualise and analyse tomographic data from different applications, e.g.,
X-ray Computed Tomography, X-ray Laminography, Neutron tomography (NT) and Positron Emission Tomography (PET).

\underline{\emph{Outline of the paper}:} In the first section we give a brief overview of CIL and introduce notation for the optimisation framework necessary for the reader to make the transition from mathematical formulation to code. Then, we consider a simple exemplar case study involving two simple 3-channel imaging tasks, i.e. colour denoising and inpainting. For the first task, the aim is to solve the Total Variation (TV) denoising problem using the Fast Gradient Projection (FGP) algorithm, \cite{BeckTeboulle}. For the inpainting problem we use the Total Generalised Variation (TGV) regularisation and solve it using the Primal-Dual Hybrid Gradient algorithm (PDHG), \cite{ChambollePock}. In the following sections we consider two real tomography applications, namely dynamic X-ray and hyperspectral CT. In section 
\ref{dynamic_section} we focus on dynamic tomographic imaging with severely limited projection data. We compare different regularisers defined over the spatiotemporal domain and under different undersampled acquisition geometries, including Tikhonov, Total variation and Directional total variation (dTV) regularisers. In the final example we deal with 4D hyperspectral tomographic data. We use a stochastic version of the PDHG, \cite{SPDHG}, with TV regularisation to reconstruct the data with different coupling between spatial and spectral dimensions.

\section{Core Imaging Library}

\subsection{Overview}

In \emph{Core Imaging Library - Part I}, \cite{CIL1} we described the main building blocks of CIL:  \module{cil.framework}, \module{cil.optimisation}, \module{cil.processors}, \module{cil.io} and \module{cil.utilities}. We illustrated the basic usage of CIL data structures,  as applied to a number of X-ray CT cases with different geometries, e.g., parallel, cone and laminography and also different modalities such as NT and PET. CIL wraps a number of third-party libraries, using \module{cil.plugins}, to perform various operations required for CT reconstruction. For instance, we can use the Astra-Toolbox \cite{vanAarle2016} or TIGRE \cite{Biguri_2016}, to perform forward and backward projection steps, filtered back projection (FBP) and Feldkamp-Davis-Kress (FDK) reconstructions for different acquisition geometries and can use the CCPi-Regularisation Toolkit (CCPi-RGL) \cite{Kazantsev}, to employ several regularisers with a CPU/GPU hardware acceleration. In addition, CIL is designed such that the data structures of the Synergistic Image Reconstruction Framework (SIRF), \cite{Ovtchinnikov2020}, from the Collaborative Computational Project in Synergistic Reconstruction for Biomedical Imaging (CCP-SynerBi), \url{www.ccpsynerbi.ac.uk}, can be used for PET and Magnetic Resonance Imaging (MRI) reconstruction, \cite{Brown2021}. 

\subsection{Optimisation Framework}

The \textbf{cil.optimisation} framework contains three structures, namely \code{Function},\, \code{Operator} and \code{Algorithm} that formalise a generic optimisation problem for imaging applications as
\begin{equation}
u^{*} =\argmin_{u\in\mathbb{X}} f(Ku) + g(u) \equiv \argmin_{u\in \mathbb{X}} \sum_{i=0}^{n-1}f_{i}K_{i}(u) + g(u).
\label{general_form}
\end{equation}
We let $\mathbb{X}, \mathbb{Y}$ denote finite-dimensional vector spaces, $K:\mathbb{X}\rightarrow \mathbb{Y}$ a linear operator with operator norm $\|K\| = \max \{ \|Ku\|_{\mathbb{Y}}: \|u\|_{\mathbb{X}}\leq 1 \}$ and proper, convex functions $f: \mathbb{Y}\rightarrow\overline{\mathbb{R}}\, \footnote{$\overline{\mathbb{R}}:=\mathbb{R}\cup\{\infty\}$}$,  $g: \mathbb{X}\rightarrow\overline{\mathbb{R}}$. Note that in certain cases, it is convenient to decompose $\mathbb{Y}=Y_{0}\times\dots \times Y_{n-1}$, $n\geq1$ and consider a separable function $f(y) := f(y_{0},\dots,y_{n-1}) = \sum_{i=0}^{n-1}f_{i}(y_{i})$ which results to the right-side formulation in \eqref{general_form}.

In the following case studies, using different definitions for the triplet ($K$, $f$, $g$), we can express optimisation problems for several imaging tasks. For example, in denoising, we let $K$ be the identity operator, in inpainting it is a mask operator that encodes missing pixel information, while it is a projection operator for tomography.  The functions $f$, $g$ allow us to define a fidelity term, that measures the distance between the acquired data $b$ and the forward-projected reconstruction image as well as well as a regulariser, which enforces a certain regularity on $u$. If the noise follows a Gaussian distribution an appropriate choice for the fidelity term is $\| Kx - b\|_{2}^{2}$. In the case of impulse noise, the $L^{1}$ norm $\|Kx - b\|_{1}$ leads to more efficient restorations and for Poisson noise the Kullback-Leibler divergence $\int Kx - b\log Kx$ is the most suitable choice. The choice of the regulariser, e.g., Tikhonov, TV and TGV, favours minimisers of \eqref{general_form} with certain geometric features and is usually weighted by positive parameters to control the influence between data fidelity and regularisation terms.

In this paper, we mainly focus on  \emph{model-based} variational problems, meaning that the forward operator $K$ and the probability distribution of the observational noise are known a-priori. In particular for X-ray CT \cite{CIL1}, PET or MRI \cite{Brown2021} applications, noise distribution and hence the appropriate distance functions are well established, see \cite{Gravel2004}. CIL may also be employed in an ad hoc fashion if the noise type is unknown, to experiment with which norm provides the best reconstruction result empirically, as shown in \cite{Rose2015}. More specific blind noise methods are an active research beyond the current scope of CIL and we hope to expand in these directions, within a general data-driven framework in the future, see for instance \cite{Arridge2019} and references therein.

In order to find an approximate solution for minimisation problems of the \eqref{general_form} form, we use different CIL \code{Algorithm} for smooth and non-smooth objective functions such as the Conjugate Gradient Least Squares (CGLS), Simultaneous Iterative Reconstruction Technique (SIRT) and proximal type algorithms, which are extensively used in this paper, such as the FGP and the PDHG, SPDHG algorithms. In the FGP algorithm, we require that the function $g$ has a proximal operator defined as 
\begin{equation}
\mathrm{prox}_{\tau g}(u)  : = \argmin_{v}  \frac{1}{2}\| v - u \|^{2}_{2} + \tau g(v), \label{definition_proximal}
\end{equation}
which has a "simple" closed form solution or can be computed efficiently numerically. Also, we assume that $f$ is continuously differentiable and has Lipschitz continuous gradient $L$. On the other hand, in the PDHG algorithm, we allow functions $f$ and $g$ to be non-differentiable and express \eqref{general_form} into a \emph{saddle point problem},
\begin{equation}
\min_{u\in\mathbb{X}}\max_{z\in\mathbb{Y}} \scalprod{}{Ku}{z} - f^{*}(z) + g(u),
\label{saddle_problem}
\end{equation}
where $f^{*}$ denotes the convex conjugate of $f$. Under this setup, PDHG can decouple the initial problem \eqref{general_form} into two simple problems, using as before the proximal operators of $g$ and in addition the proximal operator of $f^{*}$,
\begin{equation}
\mathrm{prox}_{\tau f^{*}}(u) : = \argmin_{v}  \frac{1}{2}\| v - u \|^{2}_{2} + \tau f^{*}(v).
 \label{definition_proximal_conjugate}
\end{equation} 

\section{Case Study I: Colour image processing }
\label{color_section}

We begin our first demonstration, with a case within a colour imaging framework i.e., a vector-valued image that has just 3 channels: Red, Green and Blue. Our test data is a high resolution \emph{double rainbow}\footnote{Image can be downloaded from \url{https://github.com/TomographicImaging/CIL-Data}} image taken from a smartphone of $1194\times1353$ pixels and 3 channels, see Figure \eqref{colour_example:c}. We let $$\Omega = \{ (i,j) \,| 0\leq i< M,\, 1\leq j< N,\, M=1194,\, N=1353\}$$ be a rectangular discretised grid representing our image domain and define an RGB colour image $u$ as $$u:\Omega \rightarrow \mathbb{R}^{3},\, u = (u_{1}, u_{2}, u_{3}), $$ where $u_{k}\in\mathbb{R}^{M\times N}, k=1,2,3$ represent the red, green and blue channels. We consider the cases of 
\begin{enumerate}[(a)]
\item denoising a noisy image corrupted by simulated Gaussian noise, see Figure \eqref{colour_example:a},
\item  inpainting + denoising of a noisy image corrupted by simulated Salt \& Pepper noise with missing text information, see Figure \eqref{colour_example:d}.
\end{enumerate}

\subsection{Colour Denoising}
\label{color_section:a}

\begin{figure}[h!]
\centering
\begin{subfigure}[t]{4cm}
                \centering                                                  
                \includegraphics[width=4cm]{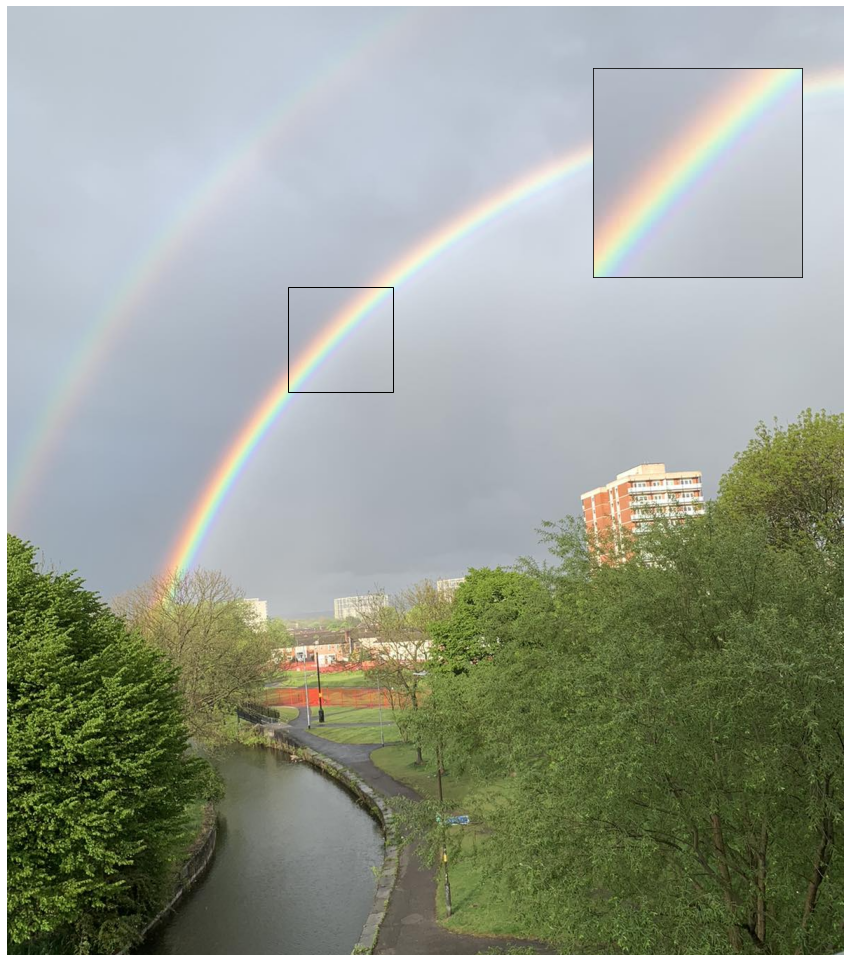}
                \caption{Ground Truth} 
                \label{colour_example:c}
\end{subfigure}
\begin{subfigure}[t]{4cm}
                \centering  
                \includegraphics[width=4cm]{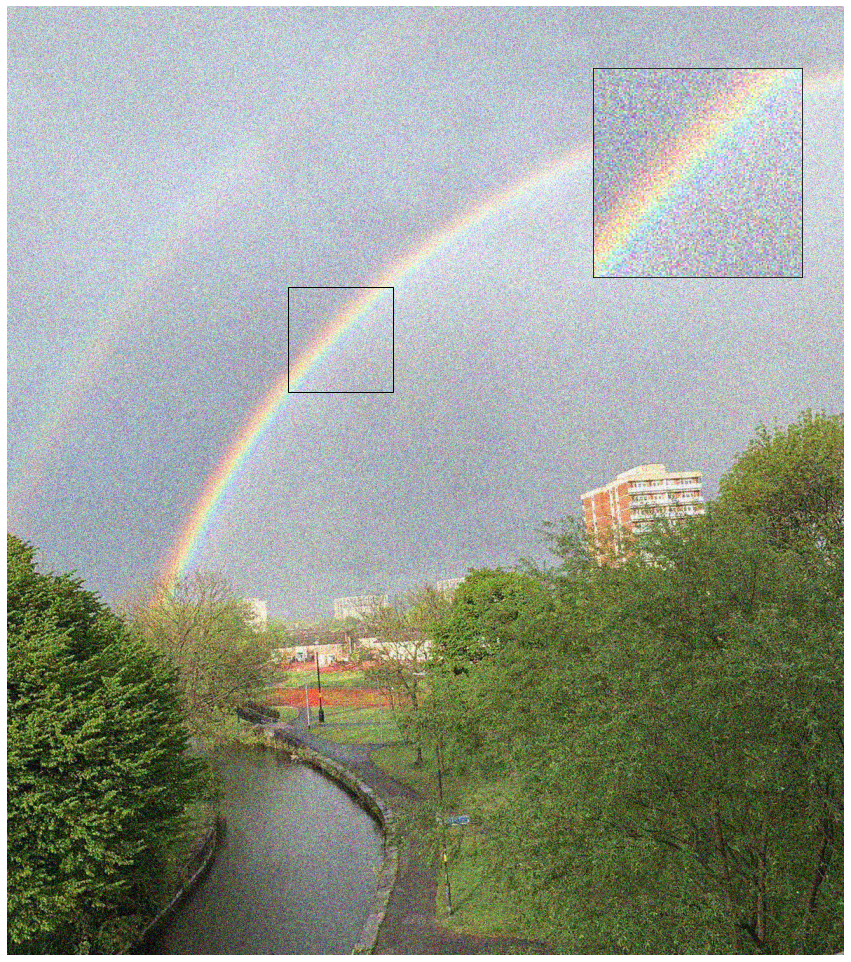}                                                
                \caption{\centering \code{noisy_data} \hspace{\textwidth} (Gaussian Noise)}
                 \label{colour_example:a}
\end{subfigure}
\begin{subfigure}[t]{4cm}
                \centering                                                  
                \includegraphics[width=4cm]{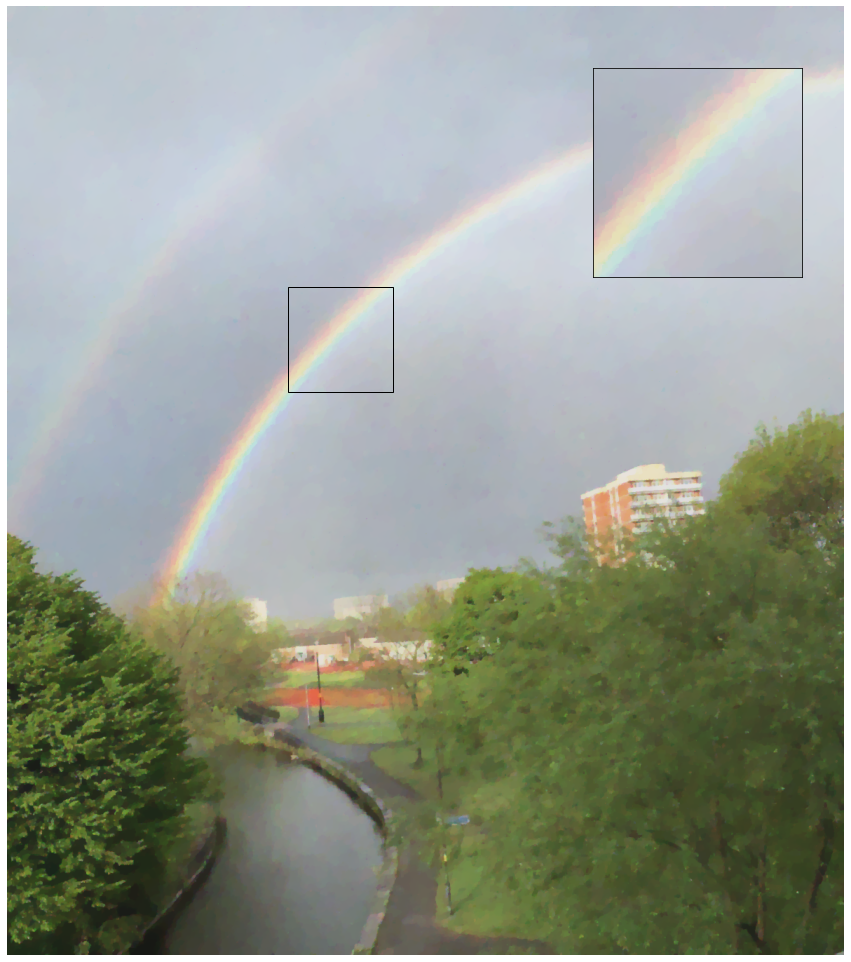}
               \caption{\centering VTV denoising ($\alpha=0.15$) \quad PSNR = 26.441, SSIM = 0.739}
                 \label{colour_example:b}
\end{subfigure}
\begin{subfigure}[t]{4cm}
                \centering        
		\includegraphics[width=4cm]{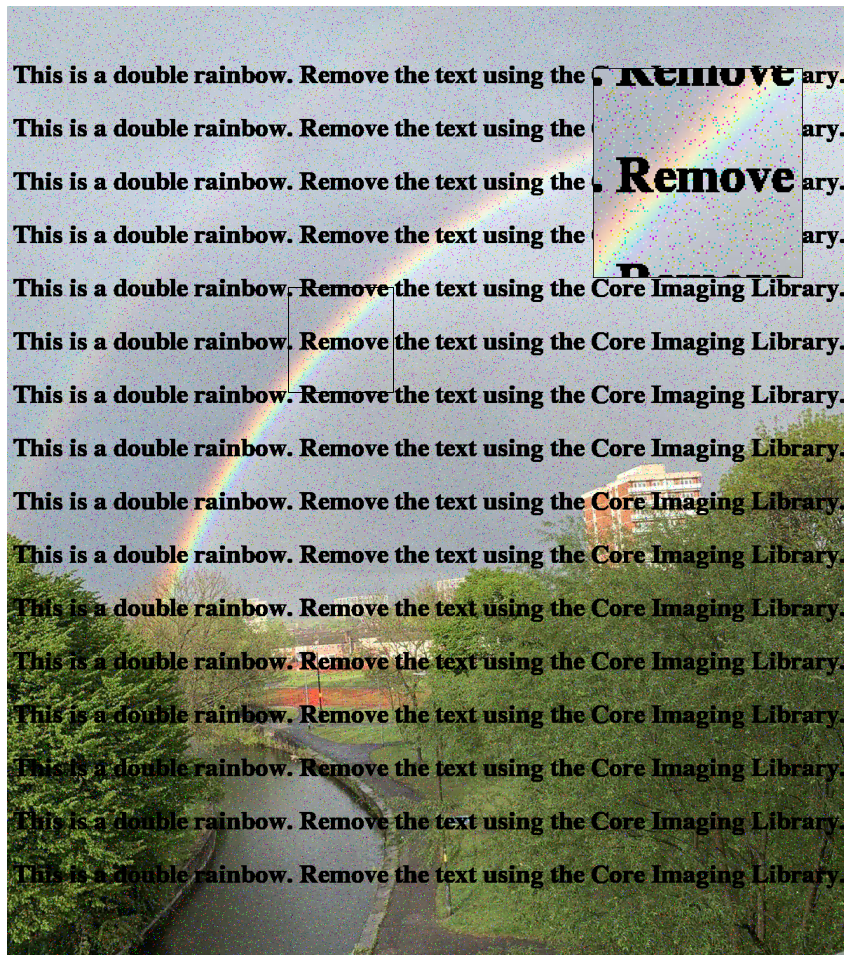}                                                          
                \caption{\centering \code{noisy_data} \hspace{\textwidth}(Salt and Pepper Noise with missing pixels)} 
                \label{colour_example:d}
\end{subfigure}
\begin{subfigure}[t]{4cm}
                \centering                                                  
                \includegraphics[width=4cm]{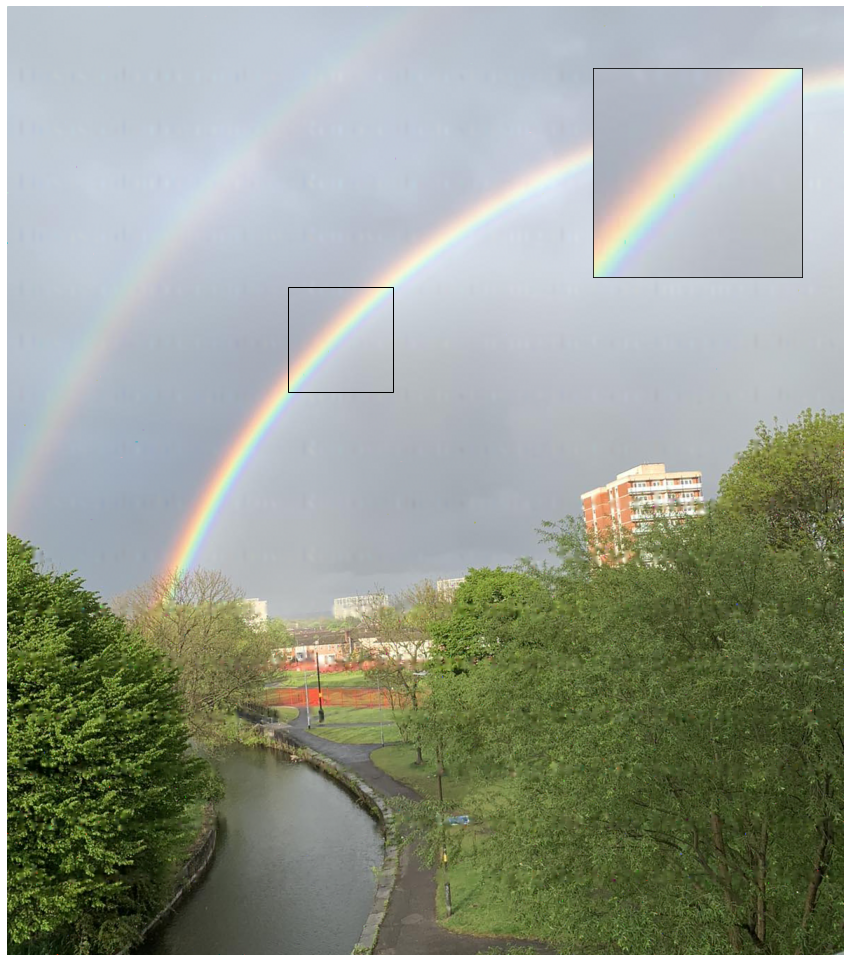}
                \caption{\centering TGV inpainting $(\alpha,\beta)= (0.5, 0.2)$  \quad\quad\quad PSNR = 32.880, SSIM = 0.960}
                \label{colour_example:e}
\end{subfigure}
\begin{subfigure}[t]{4cm}
                \centering                                                  
                \includegraphics[width=4cm]{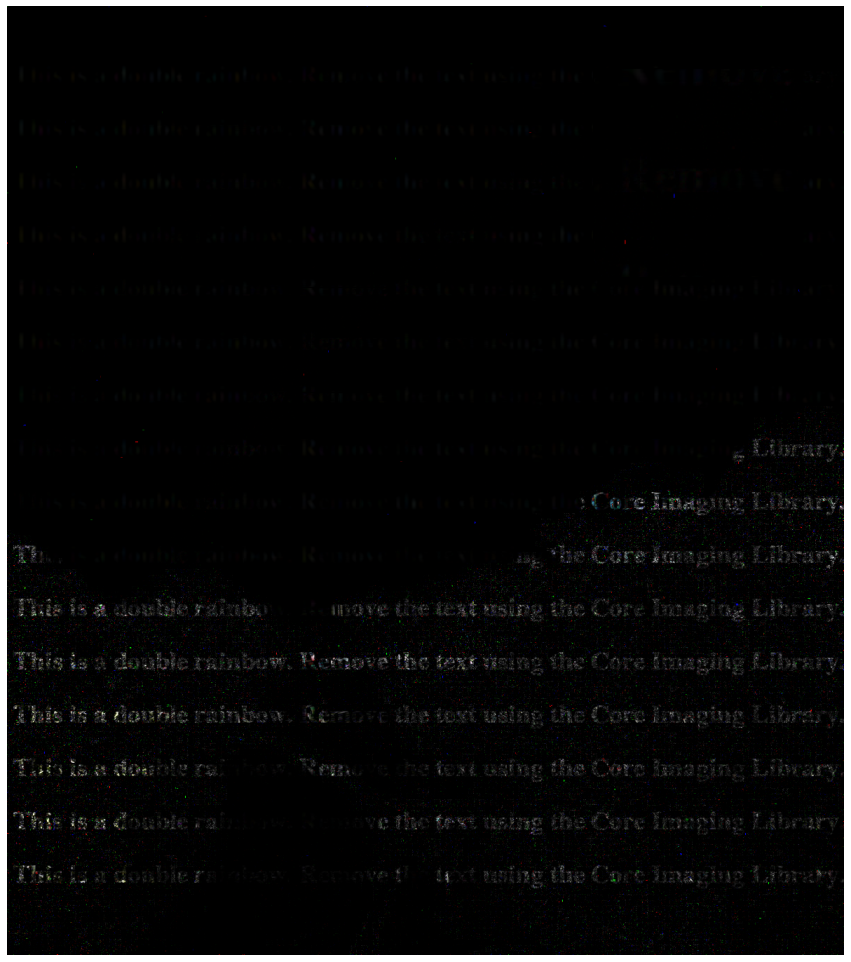}
                \caption{\centering Absolute difference \hspace{\textwidth} $|$(e) - (a)$|$ } 
                \label{colour_example:f}
\end{subfigure}
\caption{ Colour Processing: 1st row) Total Variation denoising. 2nd row) Total Generalised Variation inpainting and denoising. Regularising parameters are optimised based on the SSIM value.}
\label{tgv_example}
\end{figure}

We start with one of the most fundamental and well-studied problems in image processing, that is image denoising. In the pioneering work, Rudin, Osher and Fatemi (ROF), \cite{Rudin}, introduced the TV regulariser to tackle image denoising for grayscale images $u:\Omega\rightarrow\mathbb{R}$. Given a noisy image $b$ corrupted by additive Gaussian noise, they solve the following optimisation problem
\begin{equation}
u^{*} = \argmin_{u}  \frac{1}{2}\| b - u \|^{2}_{2} + \alpha \mathrm{TV}(u),
\label{ROF}
\end{equation}
where TV denotes the discretised total variation defined as the $\ell^{2,1}$ norm of the gradient. If the gradient is $Du:= (D_{y}u, D_{x}u)$, where $D_{y}$, $D_{x}$ denote the finite differences along the $y$ and $x$ directions respectively, we write
\begin{equation}
\mathrm{TV}(u) = \|Du\|_{2,1} = \sum_{i,j}^{M,N}\big(|(D_{y}u, D_{x}u)|_{2}\big)_{i,j} =  \sum_{i,j}^{M,N}\big(\sqrt{ (D_{y}u)^{2} + (D_{x}u)^{2}}\big)_{i,j}.
\label{tv_gray}
\end{equation}
The above definition can be extended for vector-valued images, which results in a vectorial version of the total variation. The gradient for the RGB case is now $Du=(Du_{1}, Du_{2}, Du_{3})$, where for each $k=1,2,3$, $Du_{k}:=(D_{y}u_{k}, D_{x}u_{k})$. The Vectorial Total Variation (VTV), see \cite{Duran}, is defined as
\begin{equation}
\begin{aligned}
\mathrm{VTV}(u)  := \|D u\|_{2,1}  & = \sum_{i,j}^{M,N}\big(|(Du_{1}, Du_{2}, Du_{3})|_{2}\big)_{i,j}  =  \sum_{i,j}^{M,N}\bigg(\sqrt{(|Du_{1}|_{2}^{2})_{i,j} + |Du_{2}|_{2}^{2} + |Du_{3}|_{2}^{2} }\bigg)_{i,j}\\
					      &  =  \sum_{i,j}^{M,N}\bigg(  \sqrt{  \sum_{k=1}^{3}   (D_{y} u_{k})^{2} + (D_{x} u_{k})^{2}  }     \bigg)_{i,j}
\end{aligned}
\label{tv_color}
\end{equation}
For both grayscale and coloured images, we can set up the regularisers \eqref{tv_gray} and \eqref{tv_color} using the \code{TotalVariation} function in CIL. The optimisation problem that we solve for the colour denoising is similar to \eqref{ROF} but using the VTV regulariser, i.e.,
 \begin{equation}
u^{*} = \argmin_{u}  \frac{1}{2}\| b - u \|^{2}_{2} + \alpha \mathrm{VTV}(u),
\label{VROF}
\end{equation}
where $b$ is shown in Figure \eqref{colour_example:a}. One can observe that \eqref{VROF} is in fact the proximal operator  \eqref{definition_proximal} with $\tau=1.0$ evaluated at $b$. We solve \eqref{VROF}, using the Fast Gradient Projection (FGP) algorithm that is contained in the \code{proximal} method of the \code{TotalVariation} function.
\begin{center}
\begin{tcolorbox}[
    enhanced,
    attach boxed title to top center={yshift=-2mm},
    colback=darkspringgreen!20,
    colframe=darkspringgreen,
    colbacktitle=darkspringgreen,
    title=Proximal TV Denoising,
    text width = 12.5cm,
    fonttitle=\bfseries\color{white},
    boxed title style={size=small,colframe=darkspringgreen,sharp corners},
    sharp corners,
]
\begin{minted}{python}
VTV = 0.15*TotalVariation(max_iteration=500)
proximalVTV = VTV.proximal(noisy_data, tau=1.0)
\end{minted}
\end{tcolorbox}
\end{center}
It is clear from Figure \ref{tgv_example} that noise is reduced, while preserving the edges of the image. However, total variation is known for promoting piecewise constant reconstructions leading to images with blocky structures. This is called the staircasing effect and becomes apparent in smooth regions, see for instance the area around the rainbow in Figure \eqref{colour_example:b}.

\subsection{Colour Inpainting}
\label{colorsection_b}

Given an image where a specific region is unknown, the task of image inpainting is to recover the missing region from the known part of the image. A popular application of inpainting is in art restoration, where damaged or missing areas are repainted, i.e., filled, based on the surrounding context. We let $\mathcal{D}\subset\Omega$ be a subdomain of $\Omega$, i.e., the \emph{inpainting domain}, where no data are known and missing information should be interpolated. In this example, our input image is shown in Figure \eqref{colour_example:d}, where in addition to Salt \& Pepper noise, missing pixels from a repeated text are incorporated. A suitable data fidelity term for this kind of noise distribution is the L$^{1}$ norm that acts on the domain $\Omega\setminus \mathcal{D}$.

To overcome the staircasing artifacts that TV promotes, we employ a higher-order regulariser namely the Total Generalised Variation introduced in \cite{Bredies}. We let $\alpha, \beta>0$ be regularisation parameters and define
\begin{equation}
\mathrm{TGV}_{\alpha, \beta}(u) = \min_{w} \alpha \|D u - w \|_{2,1} + \beta\|\mathcal{E}w\|_{2,1},
\label{TGV}
\end{equation}
where $\mathcal{E}$ denotes the symmetrised gradient operator defined as $\mathcal{E}w = \frac{1}{2}(D w + D w^{T})$. The optimisation problem above provides a way of balancing between the first and second derivative of an image $u$. In particular, one expects that in the neighbourhood of edges, the second derivative $D^{2}u$ is relatively ``large",  hence  a reasonable choice is to let $w=0$ in \eqref{TGV} and recover the total variation regulariser. On the other hand, $D^{2}u$ is relatively small in smooth regions of an image and $w=Du$ is a proper condition for the minimisation problem \eqref{TGV}. Under this format, edges are preserved, as in the TV regulariser, and additionally piecewise smooth structures are promoted.

The minimisation problem under the TGV regulariser and the $L^{1}$ norm as a data fidelity term is the following:
\begin{equation}
\begin{aligned}
u^{*} =\argmin_{u} & \|\mathcal{M}u-b\|_{1} + \mathrm{TGV}_{\alpha, \beta}(u) \Leftrightarrow \\
(u^{*},w^{*}) =\argmin_{u, w} &  \|\mathcal{M}u -b\|_{1} + \alpha \|D u - w \|_{2,1} + \beta\|\mathcal{E}w\|_{2,1},
\end{aligned}
\label{TGV_L1_inpainting}
\end{equation}
where the $\mathcal{M}$ is a diagonal operator with 1 in the diagonal elements corresponding to pixels in $\Omega\setminus\mathcal{D}$ and 0 in $\mathcal{D}$. In CIL, we use the \code{MaskOperator} that accepts as an input a 2D boolean array, i.e., \code{mask}. Since we have a colour image, we employ the \code{ChannelwiseOperator} to encode the effect of missing pixels to the RGB channels. In order to solve \eqref{TGV_L1_inpainting}, we use the PDHG algorithm, where the first step is to express \eqref{TGV_L1_inpainting} in the general form of \eqref{general_form}. Let $\bm{u} = (u, w)\in \mathbb{X}$ and define an operator $K:\mathbb{X}\rightarrow\mathbb{Y}$ as
\begin{equation}
K = 
\begin{bmatrix}
\mathcal{M} & \mathcal{O}\\
D & -\mathcal{I}\\
\mathcal{O} & \mathcal{E}
\end{bmatrix} \quad\Rightarrow\quad
K\bm{u} = 
K \begin{bmatrix}
u\\
w
\end{bmatrix}=
\begin{bmatrix}
\mathcal{M}u\\
Du - w\\
\mathcal{E}w
\end{bmatrix} = 
\begin{bmatrix}
y_{1}\\
y_{2}\\
y_{3}
\end{bmatrix} = \bm{y}\in \mathbb{Y},
\label{def_K}
\end{equation}
where $\mathcal{O}$, $\mathcal{I}$ denote the zero and identity operators respectively. We continue with the definition of the functions $f$ and $g$. The function $f$ is a separable function that contains the three terms in \eqref{TGV_L1_inpainting} and is defined as
\begin{equation}
\begin{aligned}
& f(\bm{y})  := f(y_{1}, y_{2}, y_{3}) = f_{1}(y_{1}) +  f_{2}(y_{2})  +  f_{3}(y_{3}), \mbox{ where},\\
& f_{1}(y_{1}) :=  \| y_{1} - b\|_1,\, f_{2}(y_{2}) :=  \alpha \|y_{2}\|_{2,1},\, f_{3}(y_{3}) := \beta\|y_{3}\|_{2,1},
\end{aligned}
\label{def_f}
\end{equation} 
and $g(\bm{u}) = g(u,w) = O(u)\equiv 0 $ is the zero function. 

Using \eqref{def_K} and \eqref{def_f}, we have that
\begin{equation*}
\begin{aligned}
f(K\bm{u}) + g(\bm{u}) & = f\bigg(\begin{bmatrix}
\mathcal{M}u\\
Du - w\\
\mathcal{E}w
\end{bmatrix}\bigg) = f_{1}(\mathcal{M}u) + f_{2}(Du-w) + f_{3}(\mathcal{E}w) \\
& = \|\mathcal{M}u -b\|_{1} + \alpha \|D u - w \|_{2,1} + \beta\|\mathcal{E}w\|_{2,1},
\end{aligned}
\end{equation*}
which is exactly the objective function in \eqref{TGV_L1_inpainting}. 

In CIL, \eqref{def_K} can be expressed easily with the \code{BlockOperator} $K$ and it is filled row-wise. The elements are the \code{GradientOperator} $D$, the \code{IdentityOperator} $\mathcal{I}$, the \code{SymmetrisedGradientOperator} $\mathcal{E}$, the \code{ChannelwiseOperator} $\mathcal{M}$ and two \code{ZeroOperator} \hspace{-1mm}$\mathcal{O}$. The separable function in \eqref{def_f} can be expressed by the \code{BlockFunction} $f$, whose elements are the \code{L1Norm}, and two \code{MixedL21Norm} functions. Finally, $g$ is the \code{ZeroFunction} function. We choose the \code{PDHG} algorithm to solve such an optimisation problem. Without any user input, CIL will by default use primal/dual stepsizes $\sigma,\tau$ with $\sigma=1.0$ and $\tau=\frac{1.0}{\sigma\|K\|^{2}}$ that satisfy $\sigma\tau\|K\|^{2}<1$ to guarantee convergence. We can monitor its convergence every \code{update_objective_interval=100} using \code{verbose=2}. However, to speed up convergence it may be necessary to change such default values.

\begin{center}
\begin{tcolorbox}[
    enhanced,
    attach boxed title to top center={yshift=-2mm},
    colback=darkspringgreen!20,
    colframe=darkspringgreen,
    colbacktitle=darkspringgreen,
    title=PDHG: TGV-L$^{1}$ Inpainting,
    text width = 12.7cm,
    fonttitle=\bfseries\color{white},
    boxed title style={size=small,colframe=darkspringgreen,sharp corners},
    sharp corners,
]
\begin{minted}{python}
K11 = ChannelwiseOperator(MaskOperator(mask), 3, dimension='append')
K21 = GradientOperator(ig)
K32 = SymmetrisedGradientOperator(K21.range)
K12 = ZeroOperator(K32.domain, ig)
K22 = -IdentityOperator(K21.range)
K31 = ZeroOperator(ig, K32.range)
K = BlockOperator(K11, K12, K21, K22, K31, K32, shape=(3,2)) 

f1 = L1Norm(b=noisy_data)
f2 = 0.5*MixedL21Norm()
f3 = 0.2*MixedL21Norm() 
f = BlockFunction(f1, f2, f3) 

g = ZeroFunction()

pdhg = PDHG(f=f, g=g, operator=K, max_iteration=1000, 
	    update_objective_interval=100)
pdhg.run(verbose=2)

\end{minted}
\end{tcolorbox}
\end{center}

The TGV reconstruction is presented in Figure \eqref{colour_example:e}, where there are no staircasing issues and most of the repeated text is eliminated. We observe that the inpainting process behaves quite well when the background is relatively smooth, e.g., sky. However, in regions with specific textures, such as trees, leaves and grass, TGV inpainting could not completely restore the missing pixels, see the absolute difference between the ground truth and reconstruction in Figure \eqref{colour_example:f}. In the above two examples, the optimal regularisation parameters are chosen to maximise the Structural Similarity Index Measure (SSIM), \cite{Wang}. In terms of TGV, it is usually sufficient to find an optimal ratio $\frac{\beta}{\alpha}$, thus reducing the number of parameters to be optimised, in order to obtain a high-quality reconstruction, \cite{Bredies,DelosReyes2016}. 

How to select the best regularisation parameter(s) is an important question and automated methods for doing this is an active research area beyond the scope of this article. This is of particular importance in real case scenarios when the comparison with the ground truth cannot guide this choice. In this paper, we choose to follow a direct grid search of parameters for simplicity. In a future release we hope to provide reconstruction algorithms with  automated parameter selection methods providing the user with an end-to-end pipeline.
We refer the reader to the following papers where a number of methods to select the regularisation parameters are presented for different imaging applications, e.g., the L-curve method \cite{Hansen1992, Calvetti2000}, the S-curve method \cite{Niinimki2016}, the Morozov Discrepancy principle \cite{Morozov1984, Bertero_2010, YouWeiWen2012,Zhang2015} and when the regularisation parameter is spatially dependent, one can use \cite{Dong2010, Langer2016, 2002.05614, Dong_2020}.

\section{Case study II:  Dynamic Tomography }
\label{dynamic_section}

\subsection{Motivation}

The focus of this section is on Dynamic CT, \cite{Bonnet2003}, where the aim is to scan a sample that undergoes some change, be it internal, such as an evolution of its composition or due to external input, like applied torque or compression, \cite{Maire_Withers2014}. These changes must be slow with respect to the time it takes to acquire a single tomogram \cite{Gajjar2018}, otherwise the reconstructions would suffer from severe motion artifacts or the quantification would be meaningless. The duration of a CT scan is determined by the time needed to acquire a sufficient number of projections  of the sample viewed from different angles with the required signal-to-noise ratio. This is determined mainly by detector performance and X-ray source intensity, and can vary from few projections per minute, as for laboratory X-ray CT scanners, to thousands of projections per second, as in the case of synchrotrons.

One way to increase the temporal resolution  is by faster scanning through undersampling, i.e. by reducing the number of acquired projections, leading to sparse tomographic views. \emph{Sparse CT} reconstruction is a highly ill-posed problem and has  received great attention lately in the tomography community, \cite{Chen2008, EmilPan, Song2007} and especially in view of Dynamic CT, \cite{Niemi2015, Bubba_2020, Wang2017, Burger_2017}. Another beneficial consequence of Sparse CT is that it allows the reduction of radiation dose to the sample, which is for instance extremely useful in medical imaging where one can reduce both the radiation dose to the patient, and the duration of the imaging \cite{Yu2009}.

In the following, we focus on different reconstruction methods for sparse dynamic CT, using an open-access dynamic dataset available from \cite{https://doi.org/10.5281/zenodo.3696817}. The aim is to demonstrate how to increase the temporal resolution, or to reduce the radiation dose of a CT scan, without sacrificing the quality of the reconstructions. After the description of the dataset and of 3 possible undersampling configurations, we demonstrate how the standard reconstruction algorithm FBP, applied separately for each time step, leads to severe streak artifacts due to the limited number of projections. We then demonstrate how to use CIL to employ iterative reconstruction algorithms with 3 different regularisation methods that incorporate prior information in the spatiotemporal domain to obtain quantitative information, to suppress the undersampling artifacts and noise on the reconstruction.  At the end of the section we compare the results obtained with all the reconstruction methods presented and demonstrate the improvements in temporal resolution and image quality enabled by a suitably chosen iterative reconstruction algorithm.

\subsection{Data information}
\underline{\textbf{Description}}: The sample was an agarose-gel phantom, \cite{2003.02841}, perfused with a liquid contrast agent in a 50 ml Falcon test tube ($\diameter$ 29mm $\times$ 115mm) The aim of this experiment was to simulate diffusion of liquids inside plant stems, which cannot withstand high radiation doses from a denser set of measurement angles.  After the agarose solidified, five intact plastic straws were made into the gel and filled with 20\% sucrose solution to guarantee the diffusion by directing osmosis to the gel body.

\noindent\underline{\textbf{Acquisition}}: Each scan was acquired in 4.5 minutes with intermissions of approximately 15 minutes between consecutive measurements. In total, the acquisition process lasted about 3 hours leading to 17 sinograms, one for each time state. In addition, pre-scan and post-scan measurements are acquired with a noticeably higher number of projections; 720 and 1600 projections respectively. The acquired sinograms are pre-processed using Lambert-Beer negative logarithm conversion.

\noindent\underline{\textbf{Dataset}}: Every measurement consists of 360 projections with 282 detectors bins obtained from a flat-panel circular-scan cone-beam microCT-scanner. Only the central slice is provided, resulting in a \emph{2D fanbeam geometry}. For this experiment, our reconstruction volume is 256x256 of 17 time frames. Additional metadata information, such as distance from source to the detector and distance from source to the origin are provided to set up the cone beam geometry.

\subsection{Dynamic Sparse CT Setup}

\begin{figure}[h!]
\hspace{0.5cm}  
\includegraphics[scale=0.34]{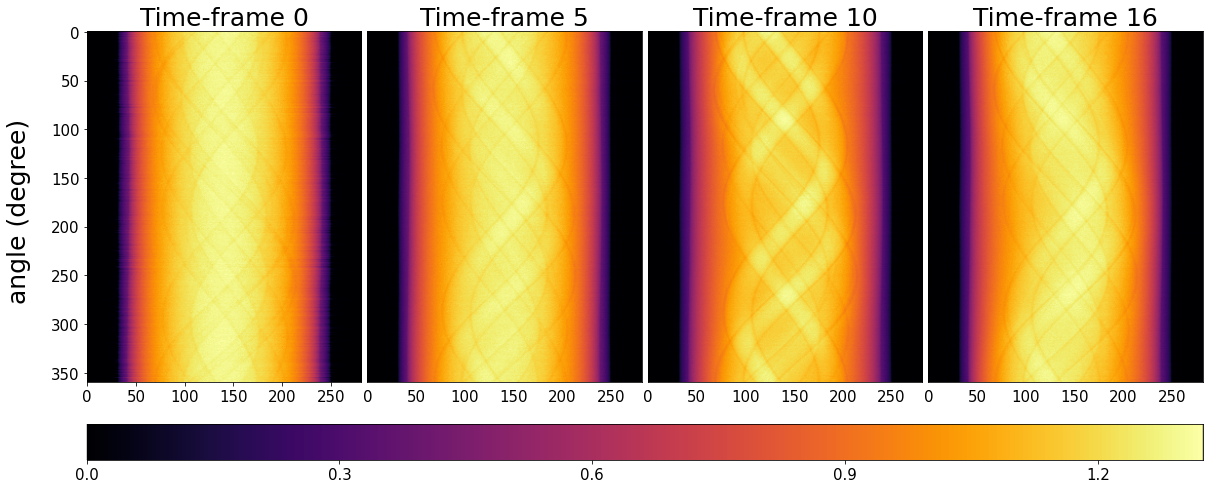}
\caption{Sinogram data with 360 projection angles of the gel-phantom for 4 different time-frames.}
\label{sinogram360_data}
\end{figure}

Firstly, we configure our \code{AcquisitionGeometry} \code{ag} for a 2D cone-beam geometry using information about the position of the source and the detector, the dimensions of the panel using \code{set_panel} and the projection angles using \code{set_angles}. In order to set up a multichannel geometry, we use \code{set_channels} which refer to the 17 time frames. Next, we allocate space for our acquisition dataset of 360 projection angles, \code{data_360}, that is filled with the corresponding sinograms for each of the 17 time frames. The Zenodo data are provided as a MATLAB mat-file that can be read in e.g., using \code{scipy.io.loadmat}, which produces a list of \code{sinograms} containing the 17 sinograms, see Figure \ref{sinogram360_data}. We obtain the default corresponding \code{ImageGeometry} \code{ig} for the acquired dataset of 360 projection angles using the \code{get_ImageGeometry} method from \code{ag}. The default dimensions are the same as the number of detector bins; here we reduce it to 256 by 256. Note that our image domain is a 2D+time spatiotemporal volume, i.e., $$\Omega = \{  (i,j,t): 0\leq i< M,\, 0\leq j< N,\, 0\leq t< T,\, M=N=256,\, T=17 \}.$$ 
\begin{center}
\begin{tcolorbox}[
    enhanced,
    attach boxed title to top center={yshift=-2mm},
    colback=darkspringgreen!20,
    colframe=darkspringgreen,
    colbacktitle=darkspringgreen,
    title=Geometries + Operator,
    text width = 12.5cm,
    fonttitle=\bfseries\color{white},
    boxed title style={size=small,colframe=darkspringgreen,sharp corners},
    sharp corners,
]
\begin{minted}{python}
ag = AcquisitionGeometry.create_Cone2D(
	source_position=[0, -100],
	detector_position=[0, 200])
        .set_panel(num_pixels=282, pixel_size=0.4)
        .set_channels(num_channels=17)
        .set_angles(angles360, angle_unit='radian')
        .set_labels(['channel', 'angle', 'horizontal'])        
        
data_360 = ag.allocate()
for i in range(ag.channels):
   data_360.fill(sinograms[i], channel=i)      
   
ig = ag.get_ImageGeometry()
ig.voxel_num_x = 256
ig.voxel_num_y = 256	      
A_360 = ProjectionOperator(ig, ag, 'gpu')      

\end{minted}
\end{tcolorbox}
\end{center}
We begin our reconstructions with the classical FBP algorithm with a Ram-Lak filter that is applied separately to the full data in each time frame, see top row in Figure \ref{dynamic_fixed_frame_diff_projs}. In a practical Sparse CT setup, 360 projection angles would not be available, however we use the full-data FBP reconstruction as our \emph{ground truth} for assessing reconstruction quality from undersampled data quantitatively and for finding the optimal regularisation parameters for the methods described below.

In order to create undersampled dynamic data for different projection angles, we employ the \code{Slicer} processor, that is described in \cite{CIL1}. We create different equi-angular undersampling patterns, using the step sizes of $[5, 10, 20]$, along the \code{angle} direction, leading to the datasets \code{data_72}, \code{data_36}, \code{data_18} of 72, 36 and 18 projections respectively. This means that we are able to increase the temporal resolution or reduce the radiation dose by a factor 5, 10 and 20 respectively. Finally, new projection operators are defined, e.g., \code{A_72},  \code{A_36},  \code{A_18}, based on the undersampled acquisition geometries and the image geometry that remains the same for all cases.

\begin{center}
\begin{tcolorbox}[
    enhanced,
    attach boxed title to top center={yshift=-2mm},
    colback=darkspringgreen!20,
    colframe=darkspringgreen,
    colbacktitle=darkspringgreen,
    title=Sparse Data + Operator (18 projections),
    text width = 12.5cm,
    fonttitle=\bfseries\color{white},
    boxed title style={size=small,colframe=darkspringgreen,sharp corners},
    sharp corners,
]
\begin{minted}{python}
data_18 = Slicer(roi={'angle':(0,360,20)})(data_360)
ag_18 = data_18.geometry
A_18 = ProjectionOperator(ig, ag_18)

\end{minted}
\end{tcolorbox}
\end{center}

\subsection{Tikhonov regularisation}
\label{dynamic_section:b}

In \cite{CIL1}, we presented in detail how one can set up Tikhonov regularisation for single channel X-ray tomography. For our current dynamic case we can formulate the identical problem, namely
\begin{align}
u^{*} = \argmin_{u} & \frac{1}{2}\| A u - b \|^{2}_{2} + \alpha \| L u \|_{2}^{2}, \quad \mbox{(Tikhonov)} \label{tikhonov_regularisation}
\end{align}
where, $b$ is now the multichannel sinogram, e.g., \code{data_360} or \code{data_18}, containing all 17 time frame sinograms and $A$ is now the corresponding multichannel projection operator, e.g., \code{A_360} or \code{A_18}. The second term in \eqref{tikhonov_regularisation} acts as a smooth regulariser, where the linear operator $L$, can be for example an Identity or a gradient operator $D$, acting over the multichannel image data. In the case of $L=D$, we offer the user two different modes for the \code{GradientOperator}, where finite differences are computed only along the spatial dimensions or for both the spatial and channel dimensions. Therefore, if \code{L = GradientOperator(ig,correlation)}, the derivatives across every direction in a 3D volume are considered, i.e., $Du = (D_{t}u, D_{y}u, D_{x}u)$, if \code{correlation='SpaceChannels'}, whereas if \code{correlation='Space'}, we take into account only the derivatives across the spatial dimensions, i.e., $Du = (D_{y}u, D_{x}u)$. In the algorithm comparison we demonstrate here finite differences over both space and channels, i.e.,  we use \code{correlation='SpaceChannels'}. The code snippet to set up \eqref{tikhonov_regularisation} in CIL is identical to the one presented in \cite{CIL1}, hence it is omitted here.

\subsection{Spatiotemporal  TV}
\label{subsec:Spatiotemporal_TV}
As a second regularisation method, we apply an edge-preserving prior by replacing the $L^{2}$ term in \eqref{tikhonov_regularisation}, with the total variation regulariser, which in a spatiotemporal setting can be employed either in a channelwise fashion or as here over the full spatio-temporal volume, i.e.,
\begin{equation}
\mathrm{TV}(u) = \|Du\|_{2,1} = \sum_{i,j,t}^{M, N, T}\big(\sqrt{ (D_{t}u)^{2} + (D_{y}u)^{2} + (D_{x}u)^{2}}\big)_{i,j,t}.
\label{tv_spatiotemporal}
\end{equation}
Under this isotropic coupling between space and time, the finite differences along the directions $t, y$ and $x$ are penalised equally with a single regularising parameter, promoting piecewise constant structures in the spatiotemporal volume by solving 
\begin{equation}
u^{*} = \argmin_{u\geq0}  \, \frac{1}{2}\| A u  - b \|^{2} + \alpha \, \mathrm{TV}(u), \quad \mbox{(Spatiotemporal TV)}.\label{spatiotemporal_tv}
\end{equation}
The above minimisation problem can be solved using the (explicit) PDHG algorithm, \cite{Jorgensen}, exactly as in the single-channel case as described in \cite{CIL1}, decomposing it into two subproblems, where the two proximal operators, \eqref{definition_proximal} and \eqref{definition_proximal_conjugate} have an explicit closed form solution. As in Section  \ref{color_section}\ref{colorsection_b}, $f$ is now a separable function, i.e., \code{BlockFunction}, containing the $\|\cdot\|^{2}_{2}$, for the acquisition data, and $\|\cdot\|_{2,1}$ norms. Consequently, we can express the operator $K$ as a \code{BlockOperator} containing the multichannel \code{ProjectionOperator} $A$ and the \code{GradientOperator}. Finally, to enforce a non-negativity constraint, we let $g$ be the \code{IndicatorBox} with \code{lower=0.0}. In the code snippet below, we define the triplet ($K$,  $f$, $g$) used in PDHG for the case of 18 projections.

\begin{center}
\begin{tcolorbox}[
    enhanced,
    attach boxed title to top center={yshift=-2mm},
    colback=darkspringgreen!20,
    colframe=darkspringgreen,
    colbacktitle=darkspringgreen,
    title= (Explicit) PDHG: Spatiotemporal TV ,
    text width = 12.5cm,
    fonttitle=\bfseries\color{white},
    boxed title style={size=small,colframe=darkspringgreen,sharp corners},
    sharp corners,
]

\begin{minted}{python}
Grad = GradientOperator(ig, correlation='SpaceChannels')
K = BlockOperator(A_18, Grad)  

f1 = 0.5*L2NormSquared(b=data_18)
f2 = alpha*MixedL21Norm()
f = BlockFunction(f1, f2)

g = IndicatorBox(lower=0)

pdhg = PDHG(f=f, g=g, operator=K, max_iteration=3000)    
pdhg.run(verbose=0)
\end{minted}
\end{tcolorbox}
\end{center}

\subsection{ Directional TV }
\label{subsec:DCT_dTV}

The third and final regularisation method uses a structure-based prior, namely the directional Total Variation (dTV). To apply this variational method to Sparse CT reconstruction, we adopt the framework of \emph{parallel level sets} introduced in \cite{6576903} and used for different applications such as multi-modal imaging, \cite{Ehrhardt2016, Ehrhardt2016MRI}. For example, an image from another modality, e.g., MRI, \emph{reference image}, is known a-priori and acts as additional information from which to propagate edge structures into the reconstruction process of another modality, e.g., PET. Another popular setup, is to use either both modalities or even channels in a joint reconstruction problem simultaneously, improving significantly the quality of the image, see for instance \cite{Ehrhardt_2014,7466848,Kazantsev_2018}.  In a parallel level set framework, two images $u$ and $v$ are called structural similar if $\nabla u$ is parallel to $\nabla v$, where $u$ is the image to be reconstructed, given the known reference image $v$. In this sense, we are able to encode additional information on the location or direction of edges for the $(u, v)$ pair. The dTV regulariser of the image $u$ given the reference image $v$ is defined as 

\begin{equation}
d\mathrm{TV}(u,v)  := \|D_{v}\nabla u\|_{2,1} = \sum_{i,j=1}^{M,N} \big(|D_{v}\nabla u|_{2}\big)_{i,j}\,,
\label{dTV_definition}
\end{equation}
where the weight $D_{v}$ depends on the normalised gradient $\xi_{v}$ of the reference image $v$, 
\begin{equation}
D_{v} = \mathbb{I}_{2\times2} - \xi_{v}\xi_{v}^T, \quad \xi_{v} = \frac{\nabla v}{\sqrt{\eta^{2} + |\nabla v|_{2}^{2}}}, \quad \eta>0.
\label{weight_D}
\end{equation}
The vector-field $\xi_{v}$ is able to capture structural information from the reference image depending on the edge parameter $\eta$ and determine which directions to be penalised. For instance, we have that $D_{v}\nabla = (1 - |\xi_{v}|_{2}^{2}) \nabla$, if $\nabla u || \nabla v$. Equivalently, if $|\xi_{v}|_{2}>0$, aligned gradients are favoured. On the other hand, if $\nabla u \perp \nabla v$ then $D_{v}\nabla = \nabla$. Finally, note that $0\leq |\xi_{v}|_{2} <1$, where the lower bound is attained for $|\nabla v|_{2} = 0$ (constant regions) and the upper bound when $|\nabla v|_{2}\rightarrow\infty$ (edges). 
\begin{figure}[b]                  
\centering                          
\includegraphics[scale=0.29]{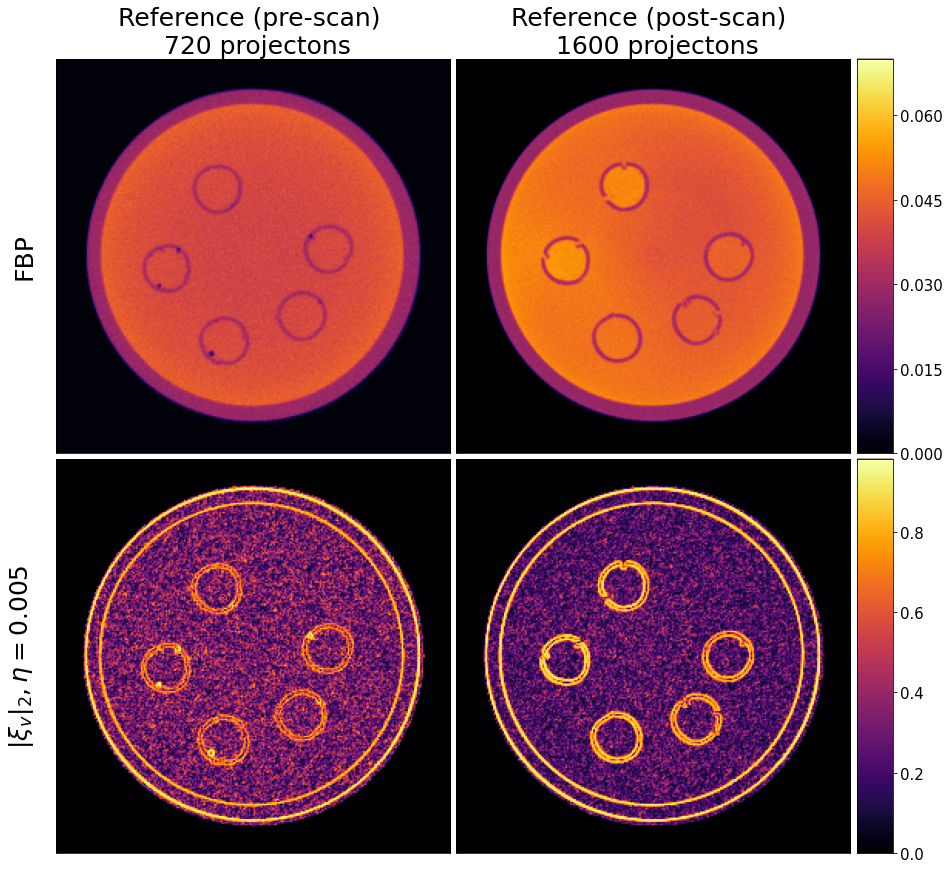}
\caption{FBP reconstructions for the acquired data (pre/post scans) with 720 and 1600 projections used as reference images for the d$\mathrm{TV}$ regulariser. The normalised gradient $\xi_{v}$ for edge parameter $\eta=0.005$. In both cases, the edges are clearly seen in the $|\xi_{v}|_2$ images. In the latter case, where noise is lower due to the larger number of projections, the edge promoting effect of dTV can be expected to be stronger.}
\label{xi_norm}
\end{figure}

In Figure \ref{xi_norm}, we show the pre- and post-scan FBP reconstructions acting as the reference images, along with $|\xi_{v}|_{2}$ that illustrates how edge information is captured by $\xi_{v}$ to be included by the dTV regularizer. For each time frame $t$, we solve the following problem
\begin{equation}
u^{*}_{t} = \argmin_{u_{t}\geq0}  \, \frac{1}{2}\| A_\text{sc} u_{t}  - b_{t} \|^{2} + \alpha \, d\mathrm{TV}(u_{t}, v_{t})\quad \mbox{(Dynamic dTV)}, \label{dynamic_dtv_problem}
\end{equation}
where $A_\text{sc}$, $b_{t}$, $u^{*}_{t}$, denote the single channel \code{ProjectionOperator}, the sinogram data and the reconstructed image for the time frame $t$ respectively. In terms of the reference images $(v_{t})_{t=0}^{T-1}$, we use $v_{0} = v_{\text{pre\_scan}}$,  i.e., the FBP reconstruction of the pre-scan data with 720 projections, and $v_{t} = v_{\text{post\_scan}}$, $t=1,\dots,T-1$, for the FBP reconstruction for the data with 1600 projections. We follow this configuration, because we notice a slight movement of the sample at the beginning of the experiment. One could apply other configurations for the reference image in the intermediate time frames. For example, in order to reconstruct the $(t+1)$th time frame, one could use the $t$th time frame reconstruction as reference. A more sophisticated reference selection approach is applied in hyperspectral computed tomography in \cite{Kazantsev_2018}.

Similarly to \eqref{spatiotemporal_tv}, we solve \eqref{dynamic_dtv_problem} using the PDHG algorithm, but with an alternative setup for the triplet ($K$,  $f$, $g$). This time one of the subproblems is not solved explicitly but an inner iterative solver is used, this is known as implicit PDHG, see \cite{Anthoine}. In particular, we let $g$ be the \code{FGP_dTV} regularising \code{Function} from \textbf{cil.plugins.ccpi\_regularisation} module of CIL, which wraps a GPU-accelerated implementation of the FGP algorithm in the CCPi-RGL toolkit. Since each time frame is solved independently of the others, the operator $K$ is now the projection operator for a single channel \code{K_sc} and the functions $f$ and $f_{0}$ are $\|\cdot\|_{2}^{2}$ norms. To store the 2D reconstruction for every time frame, we use the variable \code{solution} allocating space from the all-channel image geometry \code{ig}. Then, we use the \code{fill} method to store the reconstruction of every time step to the \code{solution}.

\begin{center}
\begin{tcolorbox}[
    enhanced,
    attach boxed title to top center={yshift=-2mm},
    colback=darkspringgreen!20,
    colframe=darkspringgreen,
    colbacktitle=darkspringgreen,
    title= (Implicit) PDHG: Dynamic dTV ,
    text width = 12.5cm,
    fonttitle=\bfseries\color{white},
    boxed title style={size=small,colframe=darkspringgreen,sharp corners},
    sharp corners,
]

\begin{minted}{python}
ag_sc = data_18.geometry.subset(channel=0)
ig_sc = ig.subset(channel=0)
K_sc = ProjectionOperator(ig_sc, ag_sc, 'gpu')

solution = ig.allocate()

f0 = 0.5*L2NormSquared(b=data_18.subset(channel=0))
g0 = alpha*FGP_dTV(reference=v_pre_scan, eta=0.005, device='gpu') 
                        
pdhg0 = PDHG(f=f0, g=g0, operator=K_sc, max_iteration=1000)
pdhg0.run(verbose=0)    
solution.fill(pdhg0.solution, channel=0)                     

g = alpha*FGP_dTV(reference=v_post_scan, eta=0.005, device='gpu') 
                        
for i in range(1, ag.channels):
    f = 0.5*L2NormSquared(b=data_18.subset(channel=i))   
    pdhg = PDHG(f=f, g=g, operator=K_sc, max_iteration=1000)
    pdhg.run(verbose 0)  
    solution.fill(pdhg.solution, channel=i)   
         
\end{minted}
\end{tcolorbox}
\end{center}

For the \code{FGP_dTV} \code{Function}, we need to specify the corresponding reference image, the regularisation parameter and the smoothing parameter $\eta$ appeared in \eqref{dTV_definition}. 
The optimal parameters $\alpha$ and $\eta$ are reported in Table \ref{table_psnr_ssim}. In the above code block, we define the projection operator, the image and acquisition geometries for the case of 18 projections. Then, the functions $f_0$, $g_0$ are used in the PDHG algorithm to reconstruct the first time frame using the \code{v_pre_scan} as a reference and the functions $f$, $g$ concern all the other frames using the \code{v_post_scan} as a reference.

\subsection{Results}

\begin{figure}[t]
\centering
\includegraphics[scale=0.38]{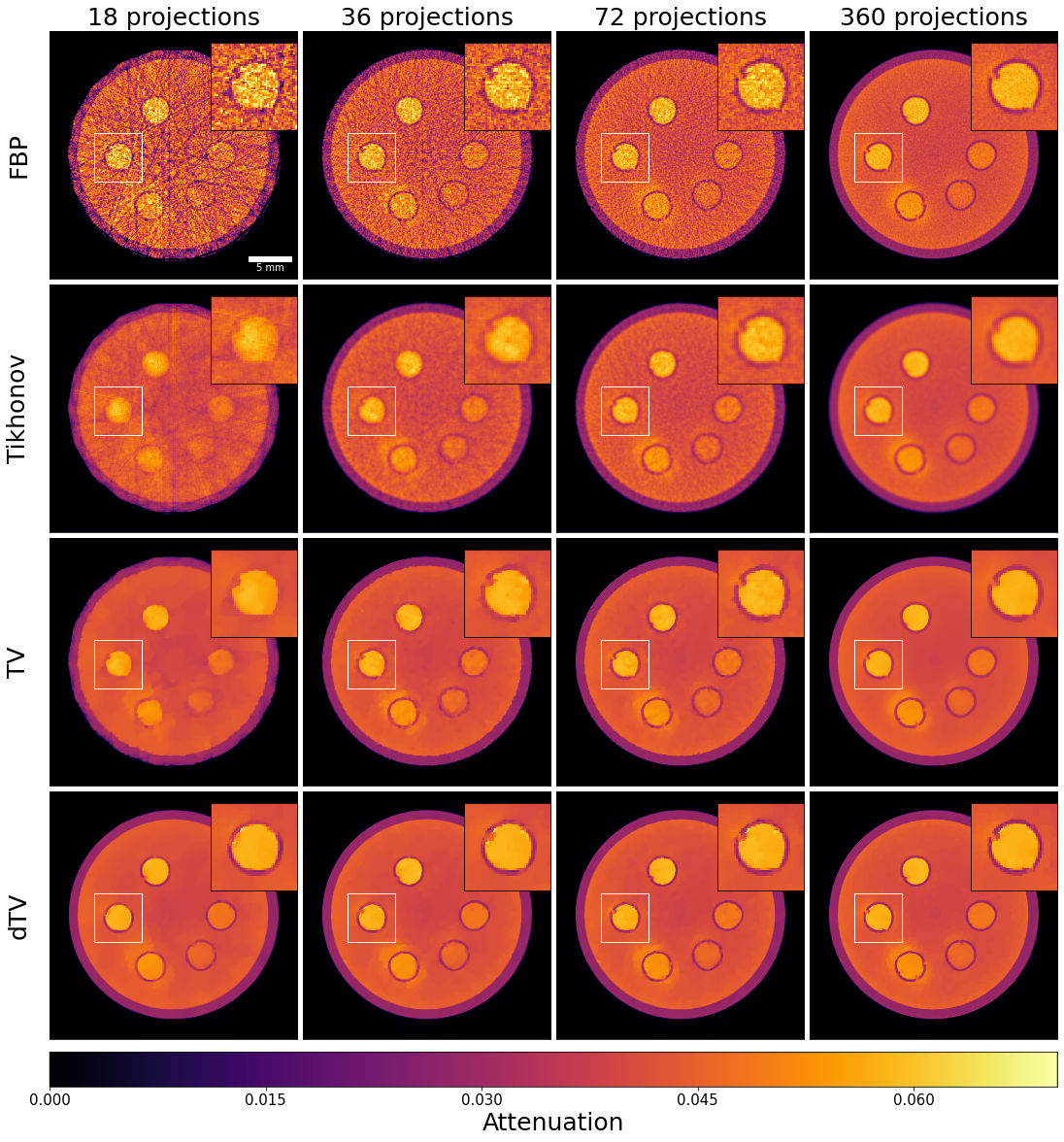}
\caption{Tomographic reconstructions of the gel phantom, for the FBP algorithm, and regularisation problems \eqref{tikhonov_regularisation}, \eqref{spatiotemporal_tv}, \eqref{dynamic_dtv_problem} with a different number of projections for the 8th time frame. The regularisation parameters and PSNR/SSIM values are reported in Table \ref{table_psnr_ssim}. All images share the same color map.}
\label{dynamic_fixed_frame_diff_projs}
\end{figure}

\begin{figure}[h!]
\centering        
\includegraphics[scale=0.47]{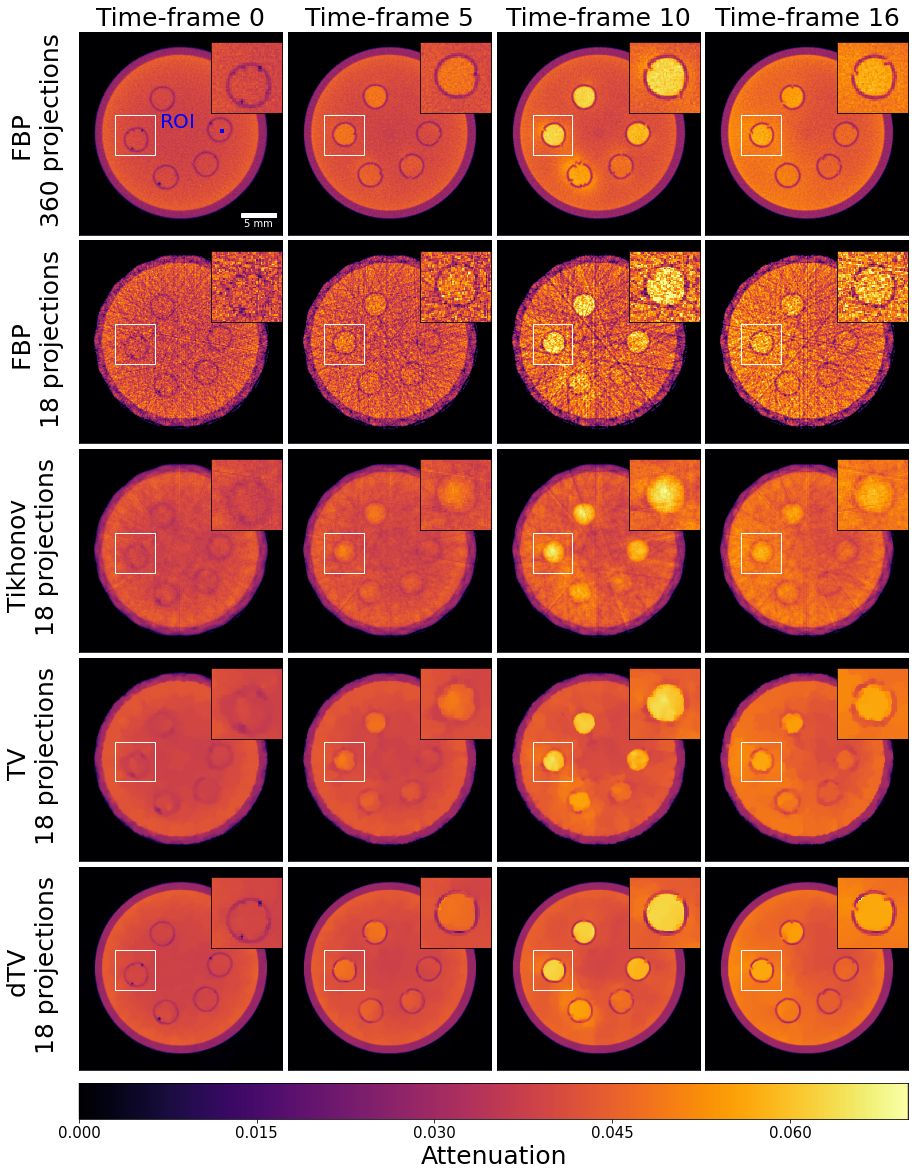}
\caption{Tomographic reconstructions of the gel phantom, for the FBP algorithm with 360 and 18 projections, and for the regularisation problems \eqref{tikhonov_regularisation}, \eqref{spatiotemporal_tv}, \eqref{dynamic_dtv_problem} with 18 projections for four different time frames. All images share the same colour map.} 
\label{dynamic_fixed_time_frame_diff_proj}
\end{figure}

In this section, we present results for all the reconstruction methods presented above, e.g., channelwise FBP algorithm, Tikhonov, TV and dTV regularisations. In Figure \ref{dynamic_fixed_frame_diff_projs}, we present a static comparison for all the reconstructions, for 3 undersampled data with 18, 36 and 72 projections as well as the full 360 projections for the 8th time frame. In Figure \ref{dynamic_fixed_time_frame_diff_proj} we provide a temporal comparison  for 4 different frames, for the most interesting case of 18 projections, as it provides the greatest reduction in acquisition time.

Although, FBP produces satisfying reconstructions for 360 projections, reducing the number of projections results in streaking artifacts and a decrease in signal-to-noise ratio which is more pronounced the higher the undersampling, see first row of Figure \ref{dynamic_fixed_frame_diff_projs}. Moreover, the quantification of the dynamic process is severely hindered by the undersampling, see for instance the first two rows of Figure \ref{dynamic_fixed_time_frame_diff_proj}, showing the FBP reconstructions for 360 and 18 projections for four different time frames. 

Compared to the FBP reconstructions, we observe that Tikhonov regularisation can suppress both the noise and the streak artifacts, especially for a very low number of projections, see second row of Figure \ref{dynamic_fixed_frame_diff_projs}. However, due to the $L^{2}$ penalty term appeared in \eqref{tikhonov_regularisation}, edges as well as small details of the image are oversmoothed. In the third row of Figure \ref{dynamic_fixed_frame_diff_projs}, we observe that noise is completely eliminated by the TV regularisation and edges are preserved around the five circular cavities. This is more obvious on the cases of 72 and 360 projections. For fewer number of projections, staircasing artifacts, introduced by TV, are more apparent both spatially and temporally, see the blocky artifacts outside the boundaries of these cavities in Figure \ref{dynamic_fixed_time_frame_diff_proj}. In addition, due to the low iodine concentration level at the earlier stage, see time frames 0 and 5 in Figure \ref{dynamic_fixed_time_frame_diff_proj}, we witness a significant loss of contrast, particularly for the case of 18 projections.

\begin{figure}[tb]
\centering                                            
\includegraphics[scale=0.28]{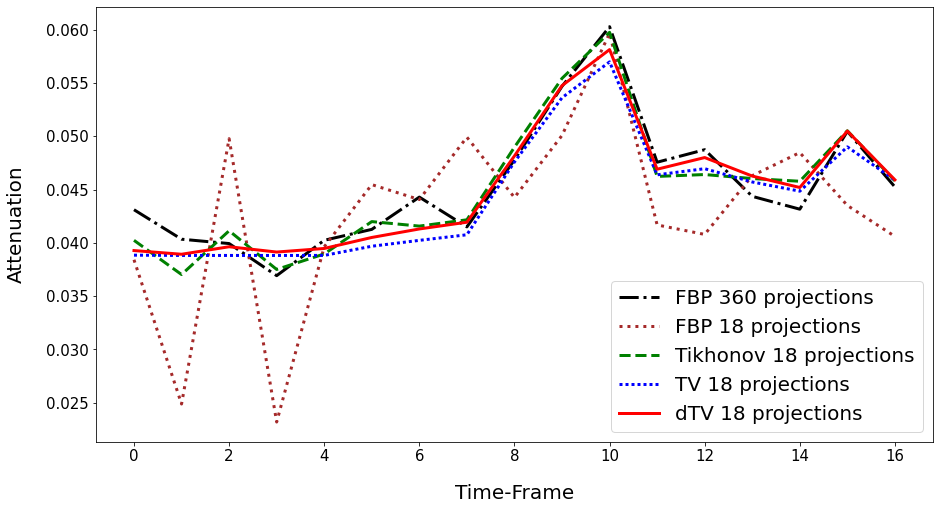}
\caption{Temporal variations in the attenuation of the gel-phantom for specific ROI (blue)
shown in the top left reconstruction of Figure 4.} 
\label{dynamic_curves}
\end{figure}

\begin{figure}[tb]
\centering                                            
\includegraphics[scale=0.28]{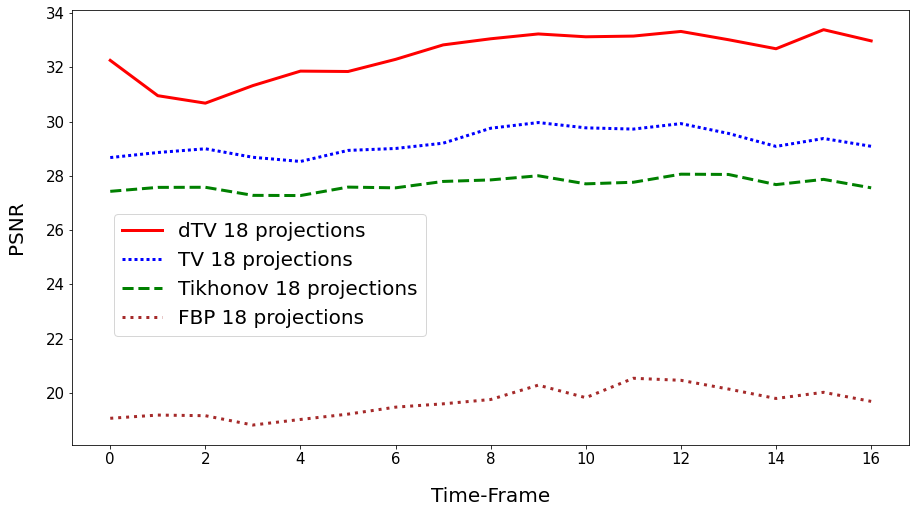}
\caption{PSNR values per time frame for the FBP, Tikhonov, TV and dTV reconstructions with 18 projections, compared with \emph{ground truth.}}
\label{psnr_dynamic}
\end{figure}

In terms of the dynamic dTV reconstructions, we observe a significant contrast improvement for all the undersampled data, see the interior of the five cavities. Furthermore, edges are now well preserved for all the time frames due to the structural information that is integrated from the reference images $v_{\text{pre\_scan}}$ and $v_{\text{post\_scan}}$. For instance, we notice sharper boundaries around the cavities compared to the TV reconstructions, specially for the lowest number of projections. This is also evident for different time frames for the 18 projections case, as one can see in the last row of Figure \ref{dynamic_fixed_time_frame_diff_proj}, where overall dTV produces the best reconstructions. In particular we note how the outer cylindrical edge is correctly reproduced as circular by dTV at 18 projections unlike all other methods which produce a polygonal outer edge due to the low number of projections.

In Figure \ref{dynamic_curves}, we compare the time activity (i.e. the reconstructed attenuation value over time) of the ground truth with the Tikhonov, TV and $d\mathrm{TV}$ reconstructions with 18 projections, for the single-pixel ROI appeared in the left image of Figure \ref{dynamic_fixed_time_frame_diff_proj}. As expected, we observe very high oscillations for the FBP reconstruction with 18 projections, which can be reduced using Tikhonov regularisation. Since there is no remarkable temporal variation until the 8th frame, we observe an almost similar behaviour for the TV and dTV reconstructions. However, dTV reconstruction provides a better contrast compared to the TV reconstruction that is more apparent after the 9th frame. Overall both dTV and TV are able to reproduce at 18 projections the single-pixel centre-of-cavity time activity at the same (or even better) quality as FBP using the full 360 projections.

In the problems \eqref{tikhonov_regularisation}, \eqref{spatiotemporal_tv} and \eqref{dynamic_dtv_problem}, all the corresponding regularisation parameters are optimised based on the highest average of PSNR values over all the time frames compared to ground truth. In Figure \ref{psnr_dynamic}, PSNR values per time frame are computed for the FBP, Tikhonov, TV and $d\mathrm{TV}$ reconstructions for 18 projections with respect to the ground truth. Overall, $d\mathrm{TV}$ has the highest PSNR for all time steps, followed by the TV and Tikhonov regularisations and finally FBP reconstruction. We also report the PSNR and SSIM values for all cases of undersampled data and the optimised parameters $\alpha$, $\eta$ for all the regularisation methods. We observe that for the very limited angular cases, e.g., (18, 36),  dTV reconstructions produce better results, whereas increasing the number of projections dTV and TV reconstructions are comparable. 

\begin{table}[tb]
\centering
\begin{tabular}{lccccccccccc}
\toprule
\textbf{Algo.}                               & \multicolumn{3}{c}{\textbf{PSNR}}& & \multicolumn{3}{c}{\textbf{SSIM}} && \multicolumn{3}{c}{\textbf{Optimal $\alpha$  ($10^{-3}$)}}  \\\cline{2-4}\cline{6-8} \cline{10-12} 
                               & 18      & 36     & 72 &  & 18      & 36     & 72   & & 18        & 36        & 72        \\ \midrule
FBP      & 19.589  & 23.452 & 27.094 & & 0.526   & 0.622  & 0.723 & & - & -& - \\ 
Tikh. & 27.157  & 30.058 & 31.523 &  & 0.647    & 0.714   & 0.761 & & 0.01 & 10 & 19 \\ 
TV     & 28.719  & 31.903 & 32.919 & & 0.720   & 0.765   & \textbf{0.784} &  & 0.72 & 0.54 & 0.81 \\ 
dTV     & \textbf{32.017}  & \textbf{32.662} & \textbf{33.083} & & \textbf{0.770}   & \textbf{0.774}  & \textbf{0.784} & & 8.1 & 7.2 & 8.1 \\ \bottomrule
\end{tabular}
\caption{PSNR and SSIM values averaged over all time frames for the FBP, Tikhonov (Tikh.), TV and dTV $(\eta=0.005)$ reconstructions at 18, 36 and 72 projections; highest scoring (in all cases dTV) in each column marked in bold. In addition the table lists optimal values of the regularisation parameter $\alpha$ (in the sense of maximising PSNR with respect to the ground truth).}
\label{table_psnr_ssim}
\end{table}

\subsection{Discussion and Conclusion}
In conclusion, we have described 3 multi-channel regularised reconstruction methods for reducing undersampling artifacts in sparse Dynamic CT and their implementation using the modular building blocks of CIL. We conducted a comparative study of algorithm performance at 3 different undersampling levels of the full dataset which simulated an increase of the temporal resolution of the acquisition by a factor 5, 10 and 20 respectively. The results demonstrated that the dTV method in particular is capable of obtaining high-quality reconstructions from reduced data and using it one can obtain the same quantitative information at a factor of 20 undersampling compared to channelwise FBP applied to full data. It is worth noting that for the dTV method of Section \ref{dynamic_section}\ref{subsec:DCT_dTV} access to high-quality pre- and post-scans is required. We want to stress that the design of the acquisition is crucial. For instance, should the experiment be acquired using a golden-ratio angular sampling scheme, \cite{Gajjar2018}, the experiment could have been let to run continuously and the time frame separation could have been decided as a post processing step. In such case, Figure \ref{dynamic_curves} could have shown a temporal resolution of up to 340=17*20 time points, with 18 angles per tomography time frame dataset.

\section{Case study III: Hyperspectral Tomography}
\label{multichannel_section}

\subsection{Motivation}

For our final case study, we focus on Hyperspectral X-ray CT imaging and how CIL can provide tools to reconstruct and analyse the internal elemental chemistry of an object. In every pixel, hyperspectral photon-counting detectors can measure the deposited energy during a certain exposure time and consequently calculate the associated photon energy of that pixel in that frame. This is repeated for many times during the scan, where finally all events are binned into a single spectrum per pixel. Moreover, these types of detectors can achieve a high energy resolution (typically less than 1 keV) and can image over hundreds of spectral bands, and allow us to distinguish materials based on the element characteristic absorption edges, i.e., K-edges.

The main goal of this study is to identify elements of gold and lead from a mineralised ore sample from a goldrich hydrothermal vein. These materials  (Au and Pb) typically appear in very low concentrations, with deposit size similar to our reconstructed voxel size. With no a priori knowledge of the distribution of these materials, and the inherently low-count hyperspectral data it is difficult to achieve satisfactory reconstruction with conventional methods. We demonstrate here how CIL can be used to easily implement a number of bespoke multi-channel regularised reconstruction methods in order to accurately identify and segment these deposits. 

In the following, we describe the dataset, then we propose and compare 3 different reconstruction methods. First we consider the standard SIRT algorithm which does not make use of any prior information, along with a variant in which the reconstruction of channel $i$ is used to warm-start reconstruction of channel $i+1$. Next we describe two advanced regularisation techniques to reconstruct 4D hyperspectral data with different correlation between spatial and energy information. Finally, we describe how the Stochastic PDHG (SPDHG) algorithm, \cite{Ehrhardt_2019,SPDHG} that uses only a subset of the whole data in every iteration, can be used to accelerate computations of large dataset reconstruction.

\subsection{Data information}

\noindent\textbf{\underline{Description:}} The sample ($\diameter$ 20mm) is extracted from a hydrothermal vein from the Leopard Mine, Silobela, Zimbabwe. It contains materials such as pyrite (FeS$_{2}$), quartz (SiO$_{2}$), gold (Au) and minor amounts of galena (PbS), chalcopyrite (CuFeS$_{2}$) and bornite (Cu$_{5}$FeS$_{4}$). 

\noindent\textbf{\underline{Acquisition:}} For the data acquisition, the authors, \cite{Egan2015}, use a colour imaging bay, in the Manchester X-Ray Imaging Facility (\href{http://www.mxif.manchester.ac.uk}{www.mxif.manchester.ac.uk}). It is designed to be a flexible work bench for spectroscopic X-ray imaging and tomography. The corresponding detector is a High Energy X-ray imaging Technology (HEXITEC) spectroscopic detector that is installed in a Nikon XTH 225 system. The sample is scanned in cone-geometry setup along 5 separate horizontal positions for increased field of view. Each sub-projection was acquired with exposure time 45 seconds (that is 225 seconds for a full stitched projection) with 120 projections covering $360^\circ$. The total scan time was 7.5 hrs. The acquired 4D raw sinogram data has three spatial dimensions and one spectral dimension, i.e., 120 projection angles of $80\times400$ pixels with 800 energy bins ranging from 1.82 keV to 186.07 keV. The reconstruction volume will be 400 by 400 by 80 voxels.

\begin{figure}[tb]
\centering           
\includegraphics[scale=0.34]{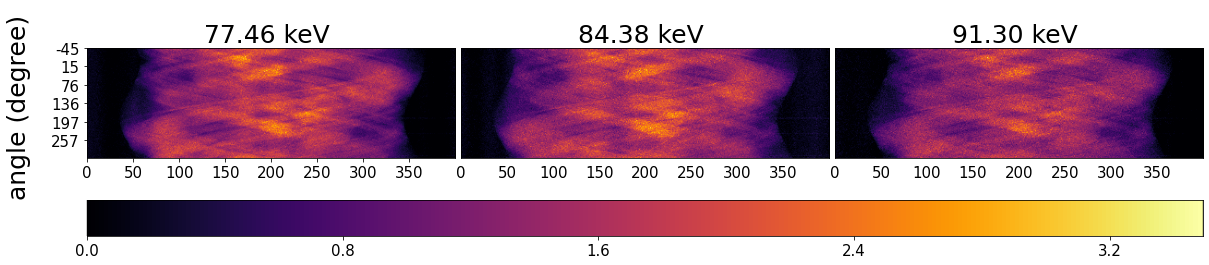}
\caption{Selected sinograms at three different energies from hyperspectral X-ray CT dataset consisting of 800 energy channels, 120 projection angles and 400 detector bins; sinogram for the 20th vertical slice out of a total 80 slices.}
\label{sinograms_hyperspectral}
\end{figure}

\noindent\textbf{\underline{Dataset:}} All the files for this study are freely available and can be downloaded from \cite{ryan_warr_2020_4157615}. It contains a)  4D hyperspectral (energy-resolved) X-ray CT projection data, b) flat-field data, c) energy in keV for every energy bin and d) geometric metadata of the cone-beam setup. Sinogram data from selected energy channels are presented in Figure \ref{sinograms_hyperspectral}. These have been pre-processed by taking the natural logarithm of the normalised intensity in every spectral band and corrected for severe vertical stripes that would cause ring artifacts, using the \code{RingRemover} processor. During the acquisition process 800 energy channels were recorded. It is possible to consider the full energy range, however for demonstration purposes we choose to examine an energy interval, [75.15 keV, 93.37 keV], of 80 channels that encompasses the K-edges of gold (Au, 80.725 keV) and lead (Pb, 88.005 keV).  

\subsection{SIRT and warm-started SIRT reconstruction}
For our first reconstruction method, we use the \code{SIRT} algorithm, an algebraic iterative method for a particular weighted least-squares problem which in addition accepts certain convex constraints such as a non-negativity constraint, see \cite{CIL1}. We enforce this as in Section \ref{dynamic_section}\ref{subsec:Spatiotemporal_TV} with the \code{IndicatorBox} function. \code{SIRT} is applied channelwise for each 3D dataset.

In addition to channelwise SIRT, one can enable a warm initialisation, that allows a basic form of channel correlation in the reconstruction as used in \cite{Egan2015}. Here, we initialise the SIRT algorithm for the $(i+1)$th channel with the solution of the $i$th channel. 


\begin{center}
\begin{tcolorbox}[
    enhanced,
    attach boxed title to top center={yshift=-2mm},
    colback=darkspringgreen!20,
    colframe=darkspringgreen,
    colbacktitle=darkspringgreen,
    title=SIRT reconstruction (warm start),
    text width = 12.5cm,
    fonttitle=\bfseries\color{white},
    boxed title style={size=small,colframe=darkspringgreen,sharp corners},
    sharp corners,
]
\begin{minted}{python}
ig = data.geometry.get_ImageGeometry()
sirt_recon4D = ig.allocate()

ag3D = data.geometry.subset(channel=0)
ig3D = ag3D.get_ImageGeometry()
A3D = ProjectionOperator(ig3D, ag3D)

sirt3D = SIRT(max_iteration=100)
x0 = ig3D.allocate()
constraint = IndicatorBox(lower=0)

for i in range(ig.channels):
    sirt3D.iteration = 0
    sirt3D.set_up(initial=x0, operator=A3D, constraint=constraint, 
    	      data=data.subset(channel=i))  
    sirt3D.run(verbose=0)
    sirt_recon4D.fill(sirt3D.solution, channel=i)
    x0.fill(sirt3D.solution)    
    
\end{minted}
\end{tcolorbox}
\end{center}

As shown in the code, we first need to extract the acquisition and image geometries, i.e., \code{ig3D} and \code{ag3D} respectively using the geometry information for one of the channels of the 4D data. Then, we set up the corresponding single-channel projection operator \code{A3D}. For every channel, we run 100 iterations of the \code{sirt3D} instance of the SIRT algorithm and its solution is filled to the corresponding channel \code{i} of the final 4D reconstruction \code{sirt_recon4D}. 

\subsection{Spatiospectral TV \& (3D + spectral)TV regularisation}

To be able to preserve edges both in the spatial domain and absorption K-edges in the energy spectrum we propose to utilise the total variation regularisation, subject to a non-negativity constraint. We first consider the total variation regulariser extended to a 4D volume, where the gradient $Du=(D_{e}u, D_{z}u, D_{y}u, D_{x}u)$ is coupled isotropically. We solve
\begin{equation}
u^{*} = \argmin_{u\geq0}  \frac{1}{2} \|Au - b\|^{2}_{2} + \alpha\|(D_{e}u, D_{z}u, D_{y}u, D_{x}u)\|_{2,1}\quad\mbox{ (Spatiospectral TV)}.\label{spatiospectral_tv}
\end{equation}
This regulariser combines together the spatial and energy variations which are penalised by a single regularising parameter. However, this may not be a good choice as the magnitude of the gradient in the spectral dimension will not necessarily be of the same order of magnitude of the one on the spatial dimensions.  Therefore it may be better to enforce separate regularisation with respect to the energy and spatial gradients, i.e., $D_{e}u$ and $(D_{z}u, D_{y}u, D_{x}u)$ respectively. We therefore consider an alternative formulation with separate decoupled TV regularisers for the spectral and spatial dimensions, i.e., 
\begin{equation}
u^{*} = \argmin_{u\geq0} \frac{1}{2} \|Au - b\|^{2}_{2} +
\beta\|D_{e}u\|_{1} + \alpha\|(D_{z}u, D_{y}u, D_{x}u)\|_{2,1} \quad\mbox{ (3D + spectral) TV}.\label{3d_energy_tv}
\end{equation}

One can solve the above problems using for instance the (explicit) PDHG algorithm. For the triplets $(K, f, g)$, the function $g$ is the same for both problems (\code{IndicatorBox}) and the difference is with respect to the operator $K$ and the functions $f$. In \eqref{spatiospectral_tv}, the operator $K$ and function $f$ are similar to the (explicit) PDHG algorithm described in the Dynamic CT section. To achieve the separation of the spatial and energy components of the \code{GradientOperator} we can set the parameter \code{split=True} so that it will split the spatial gradients and the gradient along the energy direction, i.e., $( D_{e}, (D_{z}, D_{y}, D_{x}) )$.  Similarly, we need to provide a decomposition for the function $f$, using two \code{BlockFunction} that contain the three terms presented in \eqref{spatiospectral_tv}. The following code blocks present the definition of the triplets $(K, f, g)$ for the problems \eqref{spatiospectral_tv}, \eqref{3d_energy_tv}, required to run the PDHG algorithm as described in the previous sections.

\begin{center}
\begin{tcolorbox}[
    enhanced,
    attach boxed title to top center={yshift=-2mm},
    colback=darkspringgreen!20,
    colframe=darkspringgreen,
    colbacktitle=darkspringgreen,
    title= Projection Operator + Non-negativity constraint ,
    text width = 12.5cm,
    fonttitle=\bfseries\color{white},
    boxed title style={size=small,colframe=darkspringgreen,sharp corners},
    sharp corners,
]

\begin{minted}{python}
A = ProjectionOperator(ig, data.geometry)
g = IndicatorBox(lower=0.0)


\end{minted}
\end{tcolorbox}
\end{center}

\begin{center}
\begin{tcolorbox}[
    enhanced,
    attach boxed title to top center={yshift=-2mm},
    colback=darkspringgreen!20,
    colframe=darkspringgreen,
    colbacktitle=darkspringgreen,
    title= PDHG: Spatiospectral TV,
    text width = 12.5cm,
    fonttitle=\bfseries\color{white},
    boxed title style={size=small,colframe=darkspringgreen,sharp corners},
    sharp corners,
]

\begin{minted}{python}
Grad1 = GradientOperator(ig, correlation='SpaceChannels')
K = BlockOperator(A, Grad1)
f = BlockFunction(0.5*L2NormSquared(b=data), alpha*MixedL21Norm())

\end{minted}
\end{tcolorbox}
\end{center}

\begin{center}
\begin{tcolorbox}[
    enhanced,
    attach boxed title to top center={yshift=-2mm},
    colback=darkspringgreen!20,
    colframe=darkspringgreen,
    colbacktitle=darkspringgreen,
    title= PDHG: (3D+spectral) TV,
    text width = 12.6cm,
    fonttitle=\bfseries\color{white},
    boxed title style={size=small,colframe=darkspringgreen,sharp corners},
    sharp corners,
]

\begin{minted}{python}
Grad2 = GradientOperator(ig, correlation='SpaceChannels', split=True)
K = BlockOperator(A, Grad2)
splitTV = BlockFunction(beta*L1Norm(), alpha*MixedL21Norm())
f = BlockFunction(0.5*L2NormSquared(b=data), splitTV)

\end{minted}
\end{tcolorbox}
\end{center}

\subsection{SPDHG algorithm}

Although we can follow exactly the same setup presented in the previous section to solve the above problems, one has to perform forward and backward operations of the projection operator $A$ for the whole multichannel dataset every iteration. These operations are computationally expensive, especially for large datasets. In order to overcome this problem, CIL allows the user to employ the stochastic PDHG (SPDHG) algorithm, where the above operations are applied to a randomly selected subset of the data in every iteration. SPDHG has been used for different clinical imaging applications, such as PET, \cite{Ehrhardt_2019} and motion estimation/correction in PET/MR, \cite{Brown2021}, and produces significant computational improvements over the PDHG algorithm. 

\begin{center}
\begin{minipage}{11.5cm}
\begin{algorithm}[H]
\caption{Stochastic PDHG (SPDHG)\label{alg_spdhg}}
\label{spdhg_alg}
\textbf{Inputs:} $n = \# \mathrm{subsets}$,\, $\gamma=1.0$,\, $\rho=0.99$,\,$\mathrm{probability}$ $p_{i}$, \,$i=0,\dots,n-1$\, $\\\mathrm{Stepsizes}$: $S_{i} = \gamma\frac{\rho}{\|A_{i}\|},\, T = \min\limits_{i} \frac{1}{\gamma}\frac{\rho p_{i}}{\|A_{i}\|}$, \, $i=0,\dots,n-1$\\
\textbf{Initialize:} $x^{0}, z^{0}=\overline{z}^{0}, y^{0}$ \\
\textbf{Update:} $x_{k}, y_{k}, \overline{z_{k}}$
\begin{algorithmic}[1]
\For{$k \geq 0$}
\State $x^{k+1} = \mathrm{prox}^{T}_{G}(x^{k} - T\overline{z}^{k})$
\State Select $i\in [n]$ at random with probability $p_{i}$
\State $y_{i}^{k+1} = \mathrm{prox}^{S_{i}}_{F_{i}^{*}}(y_{i}^{k} + S_{i}A_{i}x^{k+1})$
\State $\Delta z = A_{i}^{T}(y_{i}^{k+1} - y_{i}^{k})$
\State $z^{k+1} = z^{k} + \Delta z$
\State $\overline{z}^{k+1} = z^{k+1} + \frac{\Delta z}{p_{i}}$
\EndFor
\end{algorithmic}
\end{algorithm}
\end{minipage}
\end{center}

The setup of the SPDHG algorithm is similar to the PDHG algorithm with the notable differences that we need to define the subsets, which are the $n$ terms in the sum \eqref{general_form}, as well as a list of probabilities for each subset to be selected in every iteration, see Algorithm \ref{alg_spdhg} for SPDHG pseudocode. SPDHG, as well as PDHG, can be set up in an explicit form, when the regulariser is a term in the sum, or in the implicit form where it is represented by $g$.

In our hyperspectral reconstruction we use the explicit form of SPDHG. The 120 projections which constitute the acquisition data are split into $S=10$ \emph{data subsets} of 12 angles each, $b = (b_{i})_{i=0}^{S-1}$. 
In addition we will have one \emph{regulariser subset}.
We can now rewrite \eqref{spatiospectral_tv} and \eqref{3d_energy_tv}, in the form \eqref{general_form} where $n=S+1$ and 
$A_i$ represents the projection operator for a data subset, $b_i$, and with $f_i = 0.5\left\|A_iu - b_i\right\|^{2}_{2}$ for $i=0,\dots,S-1$, and $A_{n-1} = \nabla$ with $f_{n-1}$ is the \code{alpha*MixedL21Norm()} for the Spatiospectral TV regulariser or the \code{splitTV} for the (3D+spectral) TV regulariser. 

We can configure different sampling patterns, i.e. the choice of the probabilities $p_i$ to select the $i$th term of \eqref{general_form}. One may choose to assign an equal probability for every term $n$, i.e., $p_{i} = 1/n$, $i=0,\dots,n-1$; we refer to this as \emph{uniform sampling}. Another option, known as \emph{balanced sampling}, is to give a probability $0.5$ to select the regulariser subset and $0.5/S$ to select any one \emph{data subset}. In the following, we choose the balanced sampling approach and refer the reader to \cite{Ehrhardt_2019} for a  discussion on SPDHG sampling patterns.

We use the \code{Slicer} processor to obtain the data subsets $(b_{i})^{S-1}_{i=0}$. For each data subset's acquisition geometry, we create a list of operators $(A_{0}, \dots, A_{S-1})$, using the \code{ProjectionOperator} that share the same image geometry \code{ig}. For the function $f$, we use a list of the \code{L2NormSquared} \code{Function} with respect to each of the data subsets $b_{i}$ for every $i=0,\dots,S-1$. 

\begin{center}
\begin{tcolorbox}[
    enhanced,
    attach boxed title to top center={yshift=-2mm},
    colback=darkspringgreen!20,
    colframe=darkspringgreen,
    colbacktitle=darkspringgreen,
    title=SPDHG: Slice Data - Operators - Data fitting terms,
    text width = 12.7cm,
    fonttitle=\bfseries\color{white},
    boxed title style={size=small,colframe=darkspringgreen,sharp corners},
    sharp corners,
]
\begin{minted}{python}
S = 10
NA = len(data.geometry.angles)
for i in range(S):
    data_subset = Slicer(roi={'angle':(i,NA,S)})(data)
    list_A[i] = ProjectionOperator(ig, data_subset.geometry) 
    list_f[i] = 0.5*L2NormSquared(b=data_subset)
\end{minted}
\end{tcolorbox}
\end{center}
Depending on which problem we solve, the corresponding gradient operator is appended to the list of $(A_{i})$ operators, i.e., \code{Grad1} or \code{Grad2} and wrapped using the \code{BlockOperator}. Similarly, the list of data fidelity terms are wrapped as a \code{BlockFunction} $f$.

\begin{center}
\begin{tcolorbox}[
    enhanced,
    attach boxed title to top center={yshift=-2mm},
    colback=darkspringgreen!20,
    colframe=darkspringgreen,
    colbacktitle=darkspringgreen,
    title=SPDHG: Spatiospectral TV,
    text width = 12.7cm,
    fonttitle=\bfseries\color{white},
    boxed title style={size=small,colframe=darkspringgreen,sharp corners},
    sharp corners,
]
\begin{minted}{python}
K = BlockOperator(*list_A, Grad1)
f = BlockFunction(*list_f, alpha*MixedL21Norm())

\end{minted}
\end{tcolorbox}
\end{center}

\begin{center}
\begin{tcolorbox}[
    enhanced,
    attach boxed title to top center={yshift=-2mm},
    colback=darkspringgreen!20,
    colframe=darkspringgreen,
    colbacktitle=darkspringgreen,
    title=SPDHG:  (3D+spectral) TV,
    text width = 12.7cm,
    fonttitle=\bfseries\color{white},
    boxed title style={size=small,colframe=darkspringgreen,sharp corners},
    sharp corners,
]
\begin{minted}{python}
K = BlockOperator(*list_A, Grad2)
f = BlockFunction(*list_f, splitTV)
\end{minted}
\end{tcolorbox}
\end{center}
Finally, the code to set up and run the \code{SPDHG} algorithm for both \eqref{spatiospectral_tv}, \eqref{3d_energy_tv} minimisation problems under a non-negativity constraint and using a list of subset probabilities specifying balanced sampling is:

\begin{center}
\begin{tcolorbox}[
    enhanced,
    attach boxed title to top center={yshift=-2mm},
    colback=darkspringgreen!20,
    colframe=darkspringgreen,
    colbacktitle=darkspringgreen,
    title=Set up and run SPDHG algorithm,
    text width = 12.7cm,
    fonttitle=\bfseries\color{white},
    boxed title style={size=small,colframe=darkspringgreen,sharp corners},
    sharp corners,
]

\begin{minted}{python}
g = IndicatorBox(lower=0.0)	
prob = [1/(2*S)]*S + [1/2]
spdhg = SPDHG(f=f, g=g, operator=K, max_iteration=500, prob=prob)
spdhg.run()
\end{minted}
\end{tcolorbox}
\end{center}

\subsection{Results}


\begin{figure}[h!]
\centering
\includegraphics[scale=0.42]{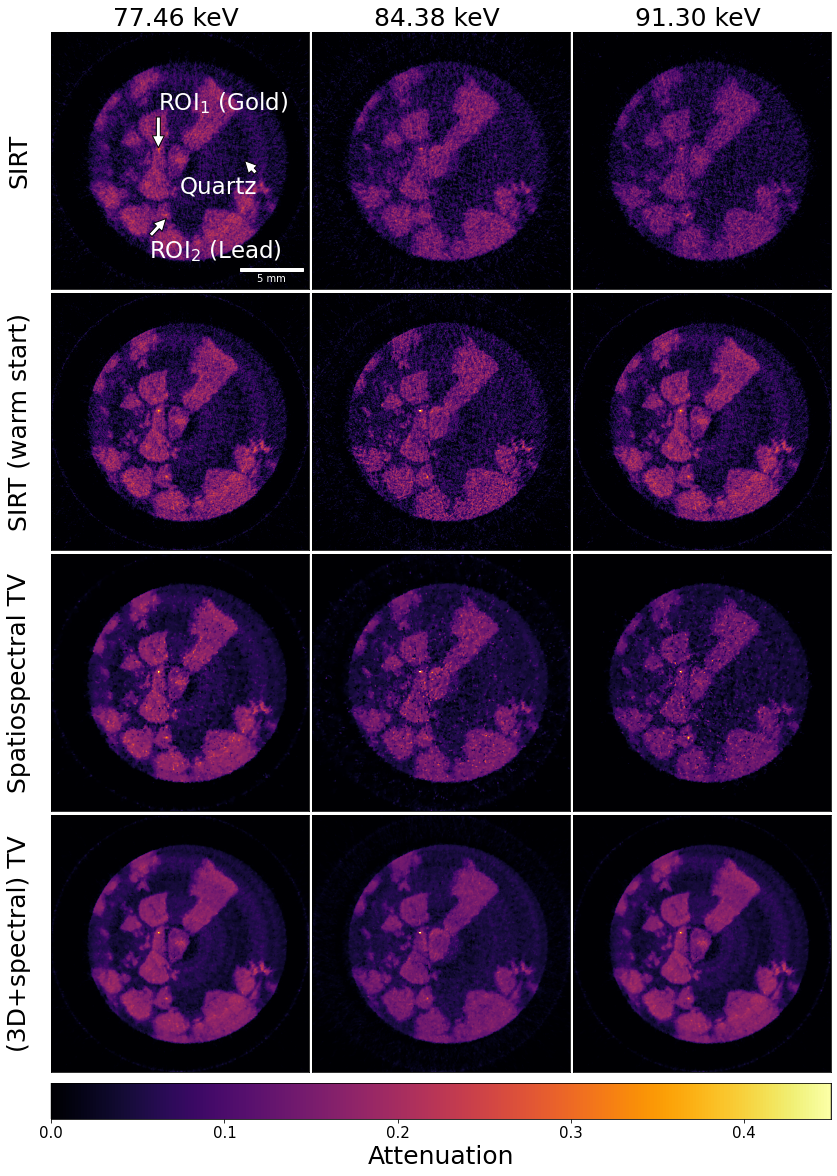}                                       
\caption{SIRT reconstructions without and with channel correlation, Spatiospectral TV and (3D+spectral) TV reconstructions using the SPDHG algorithm. Three different energies are presented at the 20th vertical slice: (1st column) below the gold K-edge, (2nd) after the gold K-edge but before the lead K-edge and (3rd) above the lead K-edge.}
\label{all_sirt_tv_reconstructions}
\end{figure}

\begin{figure}[tb]
\centering
\includegraphics[scale=0.34]{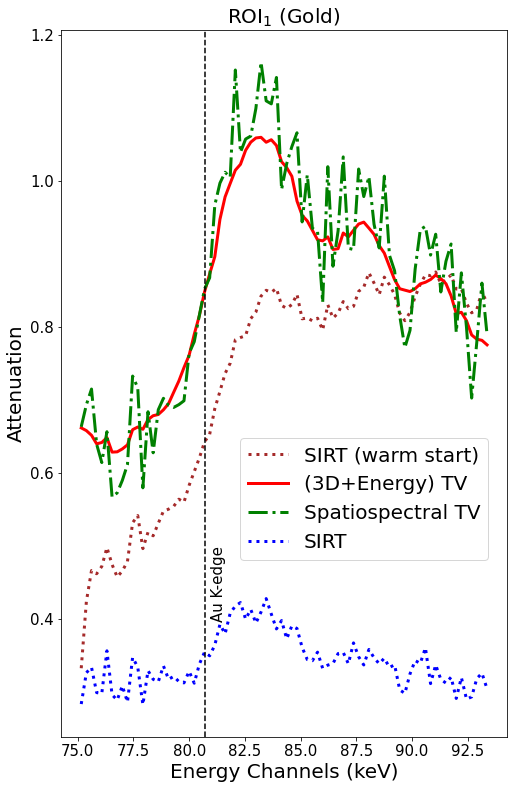}    
\includegraphics[scale=0.34]{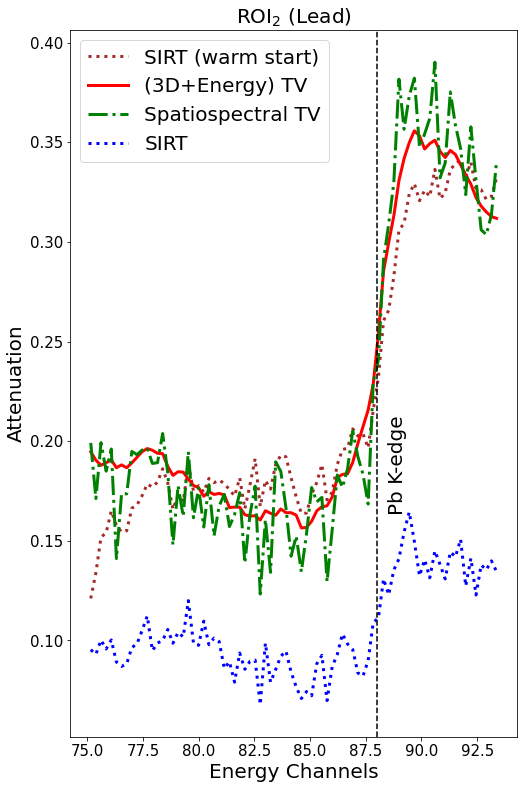}         
\caption{Attenuation plots as a function of X-ray energy for SIRT (without and with channel correlation), Spatiospectral TV and (3D+spectral) TV reconstructions for the gold (ROI$_{1}$) and lead (ROI$_{2}$) shown in Figure \ref{all_sirt_tv_reconstructions}.}
\label{energy_plots}                                   
\end{figure}

To assess performance of the algorithms considered we reconstruct the hyperspectral dataset and compare reconstructions visually (Figure \ref{all_sirt_tv_reconstructions}) and in terms of their ability to reproduce the expected sharp K-edge jumps in gold and lead containing voxels (Figure \ref{energy_plots}).

In the first rows of Figure \ref{all_sirt_tv_reconstructions}, we present results for the two versions of SIRT at 3 different energies,  below, in between and above the K-edges of Au at 80.725 keV and Pb at 88.005 keV. Since in the basic SIRT algorithm there is no regularisation to remove the noise, it is difficult to locate the ROIs of gold and lead for low energies, see the first row in Figure \ref{all_sirt_tv_reconstructions}. When the channels are linked by warm-started SIRT, we observe better contrast on these specific ROIs and both gold and lead materials are easy to distinguish. However, due to high spectral noise, the SIRT energy plots show highly oscillatory attenuation profiles, particularly for the Au plot, see Figure \ref{energy_plots}. This could make the detection of the K-edge and hence the material unreliable, particularly for cases where we do not have prior knowledge of elemental composition.


In the last rows of Figure \ref{all_sirt_tv_reconstructions}, we present the Spatiospectral TV and (3D+spectral) TV reconstructions for three different energy channels, using the SPDHG algorithm with 10 subsets and 25 epochs that correspond to 500 iterations. The regularisation parameters $\alpha$ are chosen by visual comparison reducing noise and preserving edges both spatially and along the spectral direction. Compared to the SIRT reconstructions,  noise is reduced spatially and contrast is enhanced using the Spatiospectral TV regulariser. However, moving along different channels noise is still apparent, see the first row in Figure \ref{all_sirt_tv_reconstructions} and green energy curves in Figure \ref{energy_plots}. This is because the spectral differences have less impact compared to the spatial differences in the isotropic coupling of \eqref{spatiospectral_tv}. However, the spectral noise could be reduced further by choosing a higher regularising parameter for the Spatiospectral TV, but then small features would be lost spatially due to loss of contrast. This is an inherent limitation of the coupled spatiospectral regularisation. On the contrary, in the decoupled spatial and spectral regularisation approach of the (3D+spectral) TV we have the freedom to balance the strength between space and spectral directions by suitable choices of the parameters $\alpha$ and $\beta$. In this way with the (3D+Spectral) TV we obtain higher quality reconstructions with better contrast and less noise, as seen in bottom row of Figure \ref{all_sirt_tv_reconstructions} and red energy curve of Figure \ref{energy_plots}.

\subsection{PDHG vs SPDHG}

\begin{figure}[tb]
\begin{minipage}[t]{0.48\linewidth}
\hspace{0.75cm}
\includegraphics[scale=0.33]{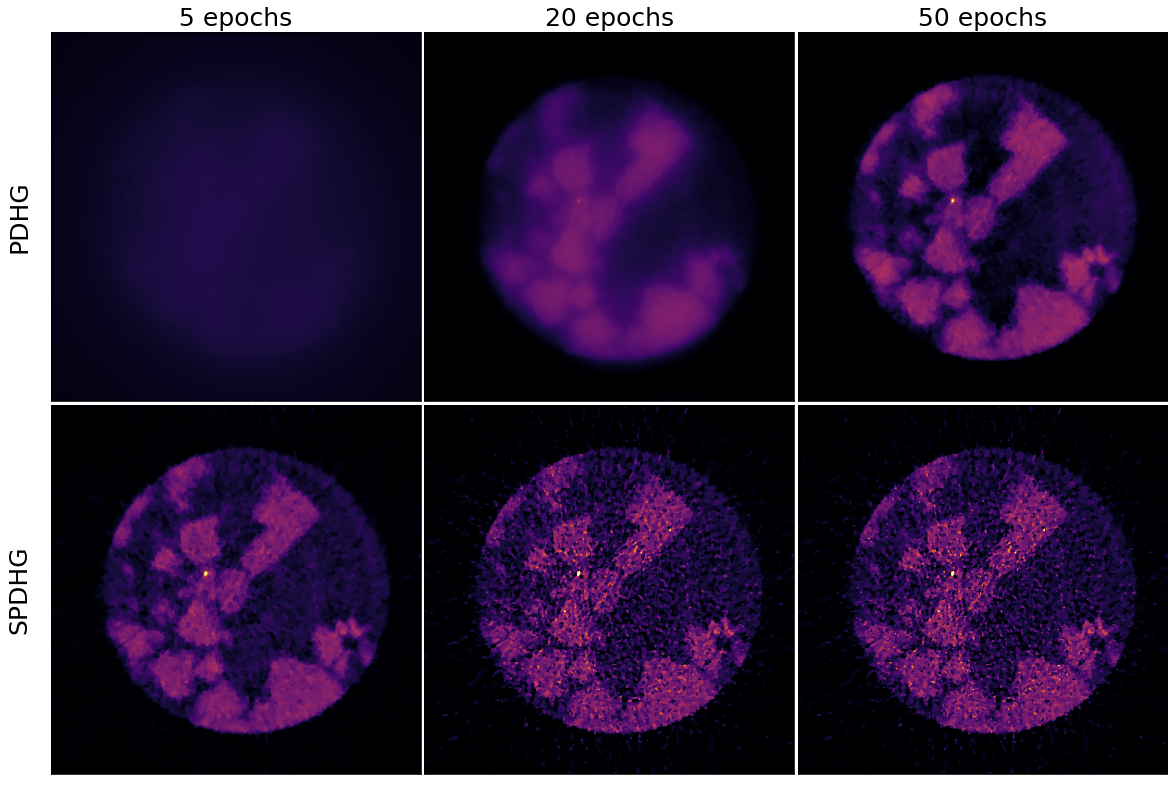}  
\end{minipage}\\[15pt]
\begin{minipage}[t]{1.1\linewidth}
\hspace{1.55cm}
\includegraphics[scale=0.6]{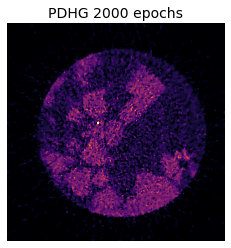}\quad\quad
\includegraphics[scale=0.24]{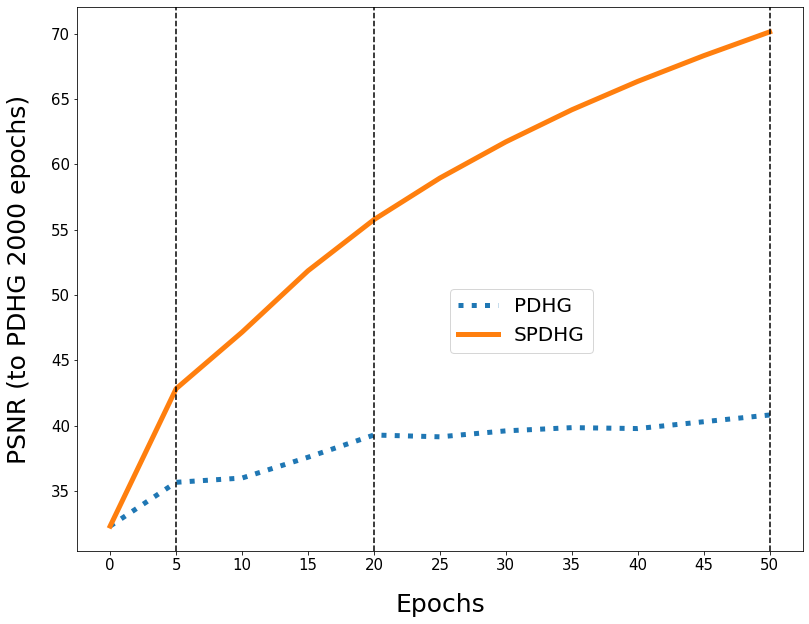}   
\end{minipage}                                           
\caption{PDHG and SPDHG reconstructions of \eqref{spatiospectral_tv} with $\alpha=0.001$ for different epochs using only 5 channels and 5 vertical slices. PSNR comparison of the PDHG and SPDHG reconstructions with respect to the PDHG reconstructions after 2000 iterations. Energy channel at 84.38 keV for the 20th vertical slice is presented.}
\label{spdhg_pdhg_plots}
\end{figure}

In Figure \ref{spdhg_pdhg_plots}, we demonstrate the computational benefit of SPDHG compared to the PDHG algorithm. On a cropped 4D dataset with only 5 channels and 5 vertical slices, we present the Spatiospectral TV reconstructions for these algorithms with respect to the number of epochs. One epoch is the expected number of iterations for the algorithm to have processed all the data, i.e., all data subsets once. For PDHG the full data is used in each iteration, so an epoch here equals an iteration.  On the other hand, for SPDHG an epoch is determined by the number of data subsets. In our case, we use $S=10$ data subsets with balanced sampling, which means that on average half the iterations call the regulariser and in the other half one of the data subsets is chosen with uniform probability. Hence, 20 iterations are required on average to process all the 10 data subsets, so an epoch for SPDHG equals 20 iterations.

We run PDHG for 2000 iterations = epochs and SPDHG for 1000 iterations = 50 epochs. We observe that even after 5 epochs, a meaningful reconstruction is obtained using the SPDHG algorithm, whereas in PDHG no structures of the rock are observed. In fact, the SPDHG reconstruction after 5 epochs is visually closer to the PDHG reconstruction after 50 epochs. This is also verified by the PSNR plot in Figure \ref{spdhg_pdhg_plots}. There, we compute for every epoch, the PSNR of the SPHDG and PDHG reconstructions against the PDHG reconstruction after 2000 epochs, which is considered as the reference image $u^{*}$, that is the converged solution of equation \eqref{tv_spatiotemporal}. 

A discussion of the computational advantages of the SPDHG vs the PDHG algorithm in tomography applications is beyond the scope of this article. Our intention was merely to demonstrate the CIL implementation of the SPDHG algorithm, which allows researchers to experiment with accelerated reconstruction of their own problems of interest. 

\subsection{Discussion and Conclusion}
In this particular rock sample, absorption K-edges are well defined and identifying the elements from these abrupt changes on the spectrum is relatively easy. However, this is not always the case, for instance if there is no prior knowledge of the sample composition, or low chemical concentration and the sample deposits are on the order of detector voxel size. In such cases K-edges may be completely concealed by the background noise of a channelwise FBP reconstruction as shown in \cite{Ryan}. Hence, we need to rely on a more sophisticated reconstruction that has the ability to suppress noise in both the spatial and spectral domains and confidently identify and quantify the elemental distribution of each material. This is particularly important when looking to perform further spectral analyses, such as the use of K-edge subtraction (KES), where we segment elemental phases based on identification of their K-edges. For a detailed task-based reconstruction quality assessment based on KES analysis, we refer the reader to \cite{Ryan}, where we compare more advanced spatiospectral reconstruction methods for a biological sample.

\section{Conclusions}
Multichannel CT imaging opens up many new possibilities in material and life sciences. Multichannel CT is intrinsically 'photon-hungry' because detected photons are shared between multiple energy or time bins. Therefore acquired tomographic datasets typically do not provide sufficient information for high quality reconstruction using traditional FBP-type algorithms. The absence of effective reconstruction methods and software capable of handling noisy and/or undersampled multichannel data hamper scientific applications of the technique.

The inverse problem framework provides methods to treat these challenging multichannel CT data through iterative reconstruction with suitable regularisation which efficiently exploits prior knowledge and inter-channel correlation. CIL implements essential building blocks, which can explicitly support multichannel reconstruction and four -- and higher -- dimensional datasets. Here, we have demonstrated the potential of CIL for multichannel CT data with three representative case-studies. Starting with a simple colour denoising and inpainting problem, we illustrated the ability to incorporate various regularisation techniques, such as classical TV, vectorial TV and TGV. We also outlined how a conventional formulation of iterative reconstruction through the optimisation framework is mapped onto CIL objects. In the second case study, we exploited reconstructions on a dynamic Sparse CT framework enforcing different prior information on spatiotemporal volume. We observed that  spatiotemporal TV is able to remove noise and streak artifacts, but due to loss of contrast, important features are lost. Using a reference image from data with dense measurements, a structural prior (dTV) is shown to enhance the reconstructions when very low number of projections are acquired. To highlight the flexibility of CIL we constructed both explicit and implicit PDHG to solve the corresponding reconstruction problems. Finally, in the last case study we endeavoured to reconstruct an energy-resolved X-ray CT dataset with high energy resolution. 
We followed the same regularisation strategy as in the second case study, i.e. a combination of edge preserving prior both in space and spectral directions, but this time we used a stochastic version of PDHG algorithm to speed-up large-scale CT reconstruction. Regularisation aided better identification of K-edges in the energy-resolved X-ray CT dataset. 

The ability to incorporate and balance various regularisation terms in the reconstruction routine is a promising approach to treat noisy and undersampled multichannel CT data, especially when using different regularisation strength for the spatial and energy domains. It is widely understood that one of the main challenges of iterative reconstruction is the high computational cost compared to traditional FBP-type methods. Although the focus of CIL is on modularity and on enabling the expression of complex optimisation problems into working code, CIL wraps hardware accelerated libraries to perform costly forward- and back-projection steps and to calculate the proximal operators of regularisation and fidelity terms, and there is a continuous effort to improve performance and resolve computational bottlenecks. In terms of supported imaging modalities, we can currently handle any tomographic modality which can be described by the Beer-Lambert law. We also provide interoperability with the Synergistic Image Reconstruction Framework (SIRF) \cite{Ovtchinnikov2020} enabling positron emission tomography and magnetic resonance imaging reconstruction using CIL. We continue to enrich the library of available algorithms, regularisers, pre- and post-processing tools along with supported imaging models and available back-ends.

\section*{Data accessibility} Python scripts to reproduce the results for all the case studies are provided at \cite{CIL2_release_software} with CIL version 21.0 that is available through Anaconda; installation instructions are at \url{https://www.ccpi.ac.uk/cil}. In addition, CIL v21.0 and subsequent releases are archived at \cite{ZenodoCIL}. The gel phantom used in Section \ref{dynamic_section} can be found in \cite{https://doi.org/10.5281/zenodo.3696817}. The rock phantom used in Section \ref{multichannel_section} can be found in \cite{ryan_warr_2020_4157615}.

\section*{Author Contribution} EPap carried out all the experiments, wrote the manuscript and developed the CIL software. JJ and EPas conceived the CIL software, assisted with the experiments and developed the CIL software. EA and GF assisted with the experiments and developed the CIL software. CD co-developed the SPDHG algorithm. RW assisted with hyperspectral case study and contributed to the CIL software. MT, WL and PW helped conceptualise and roll out the CIL software. All authors critically revised the manuscript, gave final approval for publication and agree to be held accountable for the work performed therein.

\section*{Competing interests} The author(s) declare that they have no competing interests.

\section*{Funding} This work was funded by EPSRC grants ``A Reconstruction Toolkit for
Multichannel CT'' (EP/P02226X/1), ``CCPi: Collaborative Computational Project in Tomographic Imaging'' (EP/M022498/1 and EP/T026677/1). We acknowledge the EPSRC for funding the Henry Moseley X-ray Imaging Facility through grants (EP/F007906/1, EP/F001452/1, EP/I02249X/1, EP/M010619/1, EP/F028431/1, and EP/M022498/1) which is part of the Henry Royce Institute for Advanced Materials funded by EP/R00661X/1. JSJ was partially supported by The Villum Foundation (grant no. 25893). EA was partially funded by the Federal Ministry of Education and Research (BMBF) and the Baden-W\"{u}rttemberg Ministry of Science as part of the Excellence Strategy of the German Federal and State Governments. WRBL acknowledges support from a Royal Society Wolfson Research Merit Award.  PJW and RW acknowledge support from the European Research Council grant No. 695638 CORREL-CT. CD acknowledges support from the EPSRC grant EP/S026045/1 ``PET++: Improving Localisation, Diagnosis and Quantification in Clinical and Medical PET Imaging with Randomised Optimisation''.

\section*{Acknowledgements} This work made use of computational support by CoSeC, the Computational Science Centre for Research Communities, through CCPi. Moreover, EPap would like to thank Tommi Heikkil{\"a} from the University of Helsinki for the discussions on the gel phantom. We are grateful to input from Dr. Daniil Kazantsev from the Diamond Light Source, U.K,  for early stage contributions to this work. The authors would like to thank the reviewers for their comments and efforts towards
improving this manuscript.

\bibliographystyle{vancouver}
\bibliography{final_arxiv_after_proofs_CIL2}

\begin{thebibliography}{10}

\bibitem{survey}
EPSRC X-Ray Tomography Roadmap 2018;.
\newblock
  \url{https://epsrc.ukri.org/files/research/epsrc-x-ray-tomography-roadmap-2018/}.

\bibitem{Kruger1977}
Kruger RA, Riederer SJ, Mistretta CA.
\newblock Relative properties of tomography, K-edge imaging, and K-edge
  tomography.
\newblock Medical Physics. 1977 May;4(3):244--249.
\newblock Available from: \url{https://doi.org/10.1118/1.594374}.

\bibitem{Santisteban2002}
Santisteban JR, Edwards L, Fitzpatrick ME, Steuwer A, Withers PJ, Daymond MR,
  et~al.
\newblock Strain imaging by Bragg edge neutron transmission.
\newblock Nuclear Instruments and Methods in Physics Research Section A:
  Accelerators, Spectrometers, Detectors and Associated Equipment. 2002
  Apr;481(1-3):765--768.
\newblock Available from: \url{https://doi.org/10.1016/s0168-9002(01)01256-6}.

\bibitem{ZenodoCIL}
Ametova E, Fardell G, J{\o}rgensen JS, Papoutsellis E, Pasca E. Releases of
  Core Imaging Library (CIL).
\newblock Zenodo; 2021.
\newblock Available from: \url{https://doi.org/10.5281/zenodo.4746198}.

\bibitem{Ryan}
Warr R, Ametova E, Cernik RJ, Fardell G, Handschuh S, J\o{}rgensen JS, et~al.
\newblock Enhanced hyperspectral tomography for bioimaging by spatiospectral
  reconstruction. 2021;Available from: \url{https://arxiv.org/abs/2103.04796}.

\bibitem{Evelina}
Ametova E, Burca G, Chilingaryan S, Fardell G, J{\o}rgensen JS, Papoutsellis E,
  et~al.
\newblock Crystalline phase discriminating neutron tomography using advanced
  reconstruction methods.
\newblock Journal of Physics D: Applied Physics. 2021;Available from:
  \url{https://doi.org/10.1088/1361-6463/ac02f9}.

\bibitem{BeckTeboulle}
Beck A, Teboulle M.
\newblock Fast Gradient-Based Algorithms for Constrained Total Variation Image
  Denoising and Deblurring Problems.
\newblock {IEEE} Transactions on Image Processing. 2009 Nov;18(11):2419--2434.
\newblock Available from: \url{https://doi.org/10.1109/tip.2009.2028250}.

\bibitem{ChambollePock}
Chambolle A, Pock T.
\newblock A First-Order Primal-Dual Algorithm for Convex Problems
  with~Applications to Imaging.
\newblock Journal of Mathematical Imaging and Vision. 2010 Dec;40(1):120--145.
\newblock Available from: \url{https://doi.org/10.1007/s10851-010-0251-1}.

\bibitem{SPDHG}
Chambolle A, Ehrhardt MJ, Richt{\'{a}}rik P, Sch\"{o}nlieb CB.
\newblock Stochastic Primal-Dual Hybrid Gradient Algorithm with Arbitrary
  Sampling and Imaging Applications.
\newblock {SIAM} Journal on Optimization. 2018 Jan;28(4):2783--2808.
\newblock Available from: \url{https://doi.org/10.1137/17m1134834}.

\bibitem{CIL1}
J{\o}rgensen JS, Ametova E, Burca G, Fardell G, Papoutsellis E, Pasca E, et~al.
\newblock {Core Imaging Library -- Part I: a versatile Python framework for
  tomographic imaging}.
\newblock Phil Trans R Soc A 20200192. 2021;Available from:
  \url{https://doi.org/10.1098/rsta.2020.0192}.

\bibitem{vanAarle2016}
van Aarle W, Palenstijn WJ, Cant J, Janssens E, Bleichrodt F, Dabravolski A,
  et~al.
\newblock Fast and flexible X-ray tomography using the {ASTRA} toolbox.
\newblock Optics Express. 2016 Oct;24(22):25129.
\newblock Available from: \url{https://doi.org/10.1364/oe.24.025129}.

\bibitem{Biguri_2016}
Biguri A, Dosanjh M, Hancock S, Soleimani M.
\newblock {TIGRE}: a {MATLAB}-{GPU} toolbox for {CBCT} image reconstruction.
\newblock Biomedical Physics {\&} Engineering Express. 2016 sep;2(5):055010.
\newblock Available from: \url{https://doi.org/10.1088/2057-1976/2/5/055010}.

\bibitem{Kazantsev}
Kazantsev D, Pasca E, Turner MJ, Withers PJ.
\newblock {CCPi}-Regularisation toolkit for computed tomographic image
  reconstruction with proximal splitting algorithms.
\newblock {SoftwareX}. 2019 Jan;9:317--323.
\newblock Available from: \url{https://doi.org/10.1016/j.softx.2019.04.003}.

\bibitem{Ovtchinnikov2020}
Ovtchinnikov E, Brown R, Kolbitsch C, Pasca E, da~Costa-Luis C, Gillman AG,
  et~al.
\newblock {SIRF}: Synergistic Image Reconstruction Framework.
\newblock Computer Physics Communications. 2020 Apr;249:107087.
\newblock Available from: \url{https://doi.org/10.1016/j.cpc.2019.107087}.

\bibitem{Brown2021}
Brown R, Kolbitsch C, Delplancke C, Papoutsellis E, Mayer J, Ovtchinnikov E,
  et~al.
\newblock {Motion estimation and correction for simultaneous PET/MR using SIRF
  and CIL}.
\newblock Phil Trans R Soc A 20200208. 2021;Available from:
  \url{https://doi.org/10.1098/rsta.2020.0208}.

\bibitem{Gravel2004}
Gravel P, Beaudoin G, DeGuise JA.
\newblock A Method for Modeling Noise in Medical Images.
\newblock {IEEE} Transactions on Medical Imaging. 2004 Oct;23(10):1221--1232.
\newblock Available from: \url{https://doi.org/10.1109/tmi.2004.832656}.

\bibitem{Rose2015}
Rose S, Andersen MS, Sidky EY, Pan X.
\newblock Noise properties of {CT} images reconstructed by use of constrained
  total-variation, data-discrepancy minimization.
\newblock Medical Physics. 2015 May;42(5):2690--2698.
\newblock Available from: \url{https://doi.org/10.1118/1.4914148}.

\bibitem{Arridge2019}
Arridge S, Maass P, \"{O}ktem O, Sch\"{o}nlieb CB.
\newblock Solving inverse problems using data-driven models.
\newblock Acta Numerica. 2019 May;28:1--174.
\newblock Available from: \url{https://doi.org/10.1017/s0962492919000059}.

\bibitem{Rudin}
Rudin L, Osher S, Fatemi E.
\newblock Nonlinear total variation based noise removal algorithms.
\newblock Physica D: Nonlinear Phenomena. 1992;60:259--268.
\newblock Available from: \url{http://dx.doi.org/10.1016/0167-2789(92)90242-F}.

\bibitem{Duran}
Duran J, Moeller M, Sbert C, Cremers D.
\newblock {Collaborative Total Variation: A General Framework for Vectorial TV
  Models}.
\newblock SIAM Journal on Imaging Sciences. 2016;9(1):116--151.
\newblock Available from: \url{https://doi.org/10.1137/15M102873X}.

\bibitem{Bredies}
Bredies K, Kunisch K, Pock T.
\newblock Total Generalized Variation.
\newblock {SIAM} Journal on Imaging Sciences. 2010 Jan;3(3):492--526.
\newblock Available from: \url{https://doi.org/10.1137/090769521}.

\bibitem{Wang}
Wang Z, Bovik AC, Sheikh HR, Simoncelli EP.
\newblock Image quality assessment: From error visibility to structural
  similarity.
\newblock IEEE Transactions on Image Processing. 2004;13:600--612.
\newblock Available from: \url{http://dx.doi.org/10.1109/TIP.2003.819861}.

\bibitem{DelosReyes2016}
los Reyes JCD, Sch\"{o}nlieb CB, Valkonen T.
\newblock Bilevel Parameter Learning for Higher-Order Total Variation
  Regularisation Models.
\newblock Journal of Mathematical Imaging and Vision. 2016 Jun;57(1):1--25.
\newblock Available from: \url{https://doi.org/10.1007/s10851-016-0662-8}.

\bibitem{Hansen1992}
Hansen PC.
\newblock Analysis of Discrete Ill-Posed Problems by Means of the L-Curve.
\newblock {SIAM} Review. 1992 Dec;34(4):561--580.
\newblock Available from: \url{https://doi.org/10.1137/1034115}.

\bibitem{Calvetti2000}
Calvetti D, Morigi S, Reichel L, Sgallari F.
\newblock Tikhonov regularization and the L-curve for large discrete ill-posed
  problems.
\newblock Journal of Computational and Applied Mathematics. 2000
  Nov;123(1-2):423--446.
\newblock Available from: \url{https://doi.org/10.1016/s0377-0427(00)00414-3}.

\bibitem{Niinimki2016}
Niinim\"{a}ki K, Lassas M, H\"{a}m\"{a}l\"{a}inen K, Kallonen A, Kolehmainen V,
  Niemi E, et~al.
\newblock Multiresolution Parameter Choice Method for Total Variation
  Regularized Tomography.
\newblock {SIAM} Journal on Imaging Sciences. 2016 Jan;9(3):938--974.
\newblock Available from: \url{https://doi.org/10.1137/15m1034076}.

\bibitem{Morozov1984}
Morozov VA.
\newblock Methods for Solving Incorrectly Posed Problems.
\newblock Springer New York; 1984.
\newblock Available from: \url{https://doi.org/10.1007/978-1-4612-5280-1}.

\bibitem{Bertero_2010}
Bertero M, Boccacci P, Talenti G, Zanella R, Zanni L.
\newblock A discrepancy principle for Poisson data.
\newblock Inverse Problems. 2010 aug;26(10):105004.
\newblock Available from: \url{https://doi.org/10.1088/0266-5611/26/10/105004}.

\bibitem{YouWeiWen2012}
Wen YW, Chan RH.
\newblock Parameter selection for total-variation-based image restoration using
  discrepancy principle.
\newblock {IEEE} Transactions on Image Processing. 2012 Apr;21(4):1770--1781.
\newblock Available from: \url{https://doi.org/10.1109/tip.2011.2181401}.

\bibitem{Zhang2015}
Zhang C, Zhang T, Zheng J, Li M, Lu Y, You J, et~al.
\newblock {A Model of Regularization Parameter Determination in Low-Dose X-Ray
  {CT} Reconstruction Based on Dictionary Learning}.
\newblock Computational and Mathematical Methods in Medicine. 2015;2015:1--12.
\newblock Available from: \url{https://doi.org/10.1155/2015/831790}.

\bibitem{Dong2010}
Dong Y, Hinterm\"{u}ller M, Rincon-Camacho MM.
\newblock Automated Regularization Parameter Selection in Multi-Scale Total
  Variation Models for Image Restoration.
\newblock Journal of Mathematical Imaging and Vision. 2010 Dec;40(1):82--104.
\newblock Available from: \url{https://doi.org/10.1007/s10851-010-0248-9}.

\bibitem{Langer2016}
Langer A.
\newblock Automated Parameter Selection for Total Variation Minimization in
  Image Restoration.
\newblock Journal of Mathematical Imaging and Vision. 2016 Jul;57(2):239--268.
\newblock Available from: \url{https://doi.org/10.1007/s10851-016-0676-2}.

\bibitem{2002.05614}
Hinterm\"{u}ller M, Papafitsoros K, Rautenberg CN, Sun H.
\newblock Dualization and Automatic Distributed Parameter Selection of Total
  Generalized Variation via Bilevel Optimization. 2020;Available from:
  \url{https://arxiv.org/abs/2002.05614}.

\bibitem{Dong_2020}
Dong Y, Sch\"{o}enlieb CB.
\newblock Tomographic reconstruction with spatially varying parameter
  selection.
\newblock Inverse Problems. 2020 may;36(5):054002.
\newblock Available from: \url{https://doi.org/10.1088/1361-6420/ab72d4}.

\bibitem{Bonnet2003}
Bonnet S, Koenig A, Roux S, Hugonnard P, Guillemaud R, Grangeat P.
\newblock Dynamic X-ray computed tomography.
\newblock Proceedings of the {IEEE}. 2003 Oct;91(10):1574--1587.
\newblock Available from: \url{https://doi.org/10.1109/jproc.2003.817868}.

\bibitem{Maire_Withers2014}
Maire E, Withers PJ.
\newblock Quantitative X-ray tomography.
\newblock International Materials Reviews. 2014;59(1):1--43.
\newblock Available from: \url{https://doi.org/10.1179/1743280413Y.0000000023}.

\bibitem{Gajjar2018}
Gajjar P, J{\o}rgensen JS, Godinho JRA, Johnson CG, Ramsey A, Withers PJ.
\newblock New software protocols for enabling laboratory based temporal {CT}.
\newblock Review of Scientific Instruments. 2018 Sep;89(9):093702.
\newblock Available from: \url{https://doi.org/10.1063/1.5044393}.

\bibitem{Chen2008}
Chen GH, Tang J, Leng S.
\newblock Prior image constrained compressed sensing ({PICCS}): A method to
  accurately reconstruct dynamic {CT} images from highly undersampled
  projection data sets.
\newblock Medical Physics. 2008 Jan;35(2):660--663.
\newblock Available from: \url{https://doi.org/10.1118/1.2836423}.

\bibitem{EmilPan}
Sidky EY, Kao CM, Pan X.
\newblock Accurate image reconstruction from few-views and limited-angle data
  in divergent-beam CT.
\newblock J X-ray Sci Tech. 2006;14:119--139.
\newblock Available from: \url{https://arxiv.org/abs/0904.4495}.

\bibitem{Song2007}
Song J, Liu QH, Johnson GA, Badea CT.
\newblock Sparseness prior based iterative image reconstruction for
  retrospectively gated cardiac micro-{CT}.
\newblock Medical Physics. 2007 Oct;34(11):4476--4483.
\newblock Available from: \url{https://doi.org/10.1118/1.2795830}.

\bibitem{Niemi2015}
Niemi E, Lassas M, Kallonen A, Harhanen L, H\"{a}m\"{a}l\"{a}inen K, Siltanen
  S.
\newblock Dynamic multi-source X-ray tomography using a spacetime level set
  method.
\newblock Journal of Computational Physics. 2015 Jun;291:218--237.
\newblock Available from: \url{https://doi.org/10.1016/j.jcp.2015.03.016}.

\bibitem{Bubba_2020}
Bubba TA, Heikkil\"{a} T, Help H, Huotari S, Salmon Y, Siltanen S.
\newblock Sparse dynamic tomography: a shearlet-based approach for iodine
  perfusion in plant stems.
\newblock Inverse Problems. 2020 Sep;36(9):094002.
\newblock Available from: \url{https://doi.org/10.1088/1361-6420/ab9c15}.

\bibitem{Wang2017}
Wang H, Kaestner A, Zou Y, Lu Y, Guo Z.
\newblock Sparse-view Reconstruction of Dynamic Processes by Neutron
  Tomography.
\newblock Physics Procedia. 2017;88:290--298.
\newblock Available from: \url{https://doi.org/10.1016/j.phpro.2017.06.040}.

\bibitem{Burger_2017}
Burger M, Dirks H, Frerking L, Hauptmann A, Helin T, Siltanen S.
\newblock A variational reconstruction method for undersampled dynamic x-ray
  tomography based on physical motion models.
\newblock Inverse Problems. 2017 nov;33(12):124008.
\newblock Available from: \url{https://doi.org/10.1088/1361-6420/aa99cf}.

\bibitem{Yu2009}
Yu L, Liu X, Leng S, Kofler JM, Ramirez-Giraldo JC, Qu M, et~al.
\newblock Radiation dose reduction in computed tomography: techniques and
  future perspective.
\newblock Imaging in Medicine. 2009 Oct;1(1):65--84.
\newblock Available from: \url{https://doi.org/10.2217/iim.09.5}.

\bibitem{https://doi.org/10.5281/zenodo.3696817}
Heikkil\"{a} T, Help H, Meaney A. Gel phantom data for dynamic X-ray
  tomography.
\newblock Zenodo; 2020.
\newblock Available from: \url{https://doi.org/10.5281/zenodo.3696816}.

\bibitem{2003.02841}
Heikkil\"a T, Help H, Meaney A.
\newblock Gel phantom data for dynamic X-ray tomography. 2020;Available from:
  \url{https://arxiv.org/abs/2003.02841}.

\bibitem{Jorgensen}
Sidky EY, J\o{}rgensen JS, Pan X.
\newblock Convex optimization problem prototyping for image reconstruction in
  computed tomography with the Chambolle-Pock algorithm.
\newblock Phys Med Biol. 2012;10:3065--3091.
\newblock Available from: \url{https://doi.org/10.1088/0031-9155/57/10/3065}.

\bibitem{6576903}
Ehrhardt MJ, Arridge SR.
\newblock Vector-Valued Image Processing by Parallel Level Sets.
\newblock {IEEE} Transactions on Image Processing. 2014 Jan;23(1):9--18.
\newblock Available from: \url{https://doi.org/10.1109/tip.2013.2277775}.

\bibitem{Ehrhardt2016}
Ehrhardt MJ, Markiewicz P, Liljeroth M, Barnes A, Kolehmainen V, Duncan JS,
  et~al.
\newblock {PET} Reconstruction With an Anatomical {MRI} Prior Using Parallel
  Level Sets.
\newblock {IEEE} Transactions on Medical Imaging. 2016 Sep;35(9):2189--2199.
\newblock Available from: \url{https://doi.org/10.1109/tmi.2016.2549601}.

\bibitem{Ehrhardt2016MRI}
Ehrhardt MJ, Betcke MM.
\newblock Multicontrast {MRI} Reconstruction with Structure-Guided Total
  Variation.
\newblock {SIAM} Journal on Imaging Sciences. 2016 Jan;9(3):1084--1106.
\newblock Available from: \url{https://doi.org/10.1137/15m1047325}.

\bibitem{Ehrhardt_2014}
Ehrhardt MJ, Thielemans K, Pizarro L, Atkinson D, Ourselin S, Hutton BF, et~al.
\newblock Joint reconstruction of {PET}-{MRI} by exploiting structural
  similarity.
\newblock Inverse Problems. 2014 dec;31(1):015001.
\newblock Available from: \url{https://doi.org/10.1088/0266-5611/31/1/015001}.

\bibitem{7466848}
Knoll F, Holler M, Koesters T, Otazo R, Bredies K, Sodickson DK.
\newblock Joint {MR}-{PET} Reconstruction Using a Multi-Channel Image
  Regularizer.
\newblock {IEEE} Transactions on Medical Imaging. 2017 Jan;36(1):1--16.
\newblock Available from: \url{https://doi.org/10.1109/tmi.2016.2564989}.

\bibitem{Kazantsev_2018}
Kazantsev D, J{\o}rgensen JS, Andersen MS, Lionheart WRB, Lee PD, Withers PJ.
\newblock Joint image reconstruction method with correlative multi-channel
  prior for x-ray spectral computed tomography.
\newblock Inverse Problems. 2018 apr;34(6):064001.
\newblock Available from: \url{https://doi.org/10.1088/1361-6420/aaba86}.

\bibitem{Anthoine}
Anthoine S, , Aujol JF, Boursier Y, M{\'{e}}lot C, and.
\newblock Some proximal methods for Poisson intensity {CBCT} and {PET}.
\newblock Inverse Problems {\&} Imaging. 2012;6(4):565--598.
\newblock Available from: \url{https://doi.org/10.3934/ipi.2012.6.565}.

\bibitem{Ehrhardt_2019}
Ehrhardt MJ, Markiewicz P, Sch{\"o}nlieb CB.
\newblock Faster {PET} reconstruction with non-smooth priors by randomization
  and preconditioning.
\newblock Physics in Medicine {\&} Biology. 2019 nov;64(22):225019.
\newblock Available from: \url{https://doi.org/10.1088/1361-6560/ab3d07}.

\bibitem{Egan2015}
Egan CK, S~D~M J, Wilson MD, Veale MC, Seller P, Beale AM, et~al.
\newblock {3D chemical imaging in the laboratory by hyperspectral X-ray
  computed tomography}.
\newblock Scientific Reports. 2015;5(1):15979.
\newblock Available from: \url{https://doi.org/10.1038/srep15979}.

\bibitem{ryan_warr_2020_4157615}
Warr R, Egan C, Papoutsellis E, J\o{}rgensen JS, Ametova E, Cernik R, et~al..
  {Hyperspectral X-ray CT data set of mineralised ore sample with Au and Pb
  deposits}.
\newblock Zenodo; 2020.
\newblock Available from: \url{https://doi.org/10.5281/zenodo.4157615}.

\bibitem{CIL2_release_software}
Papoutsellis E, Ametova E, Delplancke C, Fardell G, J{\o}rgensen JS, Pasca E,
  et~al.. {Code to reproduce results of "Core Imaging Library Part II:
  multichannel reconstruction for dynamic and spectral tomography"}.
\newblock Zenodo; 2021.
\newblock Available from: \url{https://doi.org/10.5281/zenodo.4744745}.

\end{thebibliography}

\end{document}